\documentclass[
16pt,
onecolumn,
superscriptaddress,
nofootinbib,
 amsmath,amssymb
]{revtex4-2}
\usepackage{subcaption}
\usepackage{float}
\usepackage[colorlinks]{hyperref}
\hypersetup{
  citecolor=red,
  linkcolor= blue,
  urlcolor=blue}
\usepackage{graphicx}% Include figure files
\usepackage{dcolumn}% Align table columns on decimal point
\usepackage{bm}% bold math

\usepackage{subcaption}

\usepackage{cancel}% to cancel terms in equations

\usepackage{amsfonts}%bold math
\usepackage{xcolor}%to color some texts
\usepackage{tcolorbox}% for boxing texts

\usepackage{caption}

\newcommand{\beqa}{\begin{eqnarray}}
\newcommand{\eeqa}{\end{eqnarray}}
\newcommand{\nn}{\nonumber}
\begin{document}
\title{ Force Dipole Interactions in Tubular Fluid Membranes }
\author{Samyak Jain}
\affiliation{Department of Physics, Indian Institute of Technology Bombay, Mumbai 400076, India}
\author{Rickmoy Samanta} \affiliation{ Department of Physics, Birla Institute of Technology and Science Pilani, Hyderabad 500078, India\\Corresponding author, Email: rickmoysamanta@gmail.com }
%\author{Samyak Jain (\begin{myhindi}सम्यक जैन)\end{myhindi}\\Department of Physics, Indian Institute of Technology Bombay, Mumbai 400076, India \\Rickmoy Samanta \begin{myhindi}(रि‌‌क्मौय सामन्ता)\end{myhindi}\\ Department of Physics, Birla Institute of Technology and Science Pilani, Hyderabad 500078, India\\Corresponding author, Email: rickmoysamanta@gmail.com }
\begin{abstract}
Abstract: We construct viscous fluid flow sourced by a force dipole embedded in a cylindrical fluid membrane, coupled to external embedding fluids. We find analytic expressions for the flow, in the limit of infinitely long and thin tubular membranes. We utilize this solution to formulate the in-plane dynamics of a pair of pusher-type dipoles along the cylinder surface. We find that a mutually perpendicular dipole pair move together along helical geodesics, thus acting as \textit{curvature checkers}, analogous to vortex dipoles. Since the cylindrical geometry breaks the in-plane rotational symmetry of the membrane, there is a difference in flows along the axial ($\hat{z}$) and transverse ($\hat{\theta}$) directions of the cylinder. This in turn leads to anisotropic hydrodynamic interaction between the dipoles and is remarkably different from  flat and spherical fluid membranes. In particular, the  flow along the compact $\hat{\theta}$ direction of the cylinder has a \textit{local} rigid rotation term (independent of the angular coordinate but decays along the axis of the cylinder). Due to this feature of the flow, we observe that the interacting dipole pair initially situated along the axial direction $\hat{z}$ exhibits an overall ``drift" along the compact angular direction $\hat{\theta}$ of the tubular fluid membrane. We find that the drift for the dipole pair increases \textit{linearly} with time. Our results are relevant for  non-equilibrium dynamics of motor proteins in tubular membranes arising in nature, as well as in-vitro experiments.
\end{abstract}
\maketitle \section{Introduction} Hydrodynamic interaction of biological motors in fluid membranes and interfaces is currently an active area of research. The fluid nature of cell membrane \cite{sn72} plays an important role in various transport and mixing processes in living cells. Such two dimensional (2D) membranes are often approximated as a thin layer of viscous fluid and the associated fluid flows are governed by Low Reynolds hydrodynamics \cite{purcell}. Moreover, the membrane fluid can exchange momentum with the
external fluids surrounding the membrane and hence  they are essentially quasi-2D in nature. The pioneering works of Saffman and Delbr\"{u}ck \cite{saff1,saff2} illustrated the low Reynolds  hydrodynamics of inclusions in such quasi-2D, flat membranes, followed by several refinements \cite{hughes,evans,lg96,staj,fischer,nlp07}. The ratio of membrane and solvent viscosity gives rise to a length scale (Saffman length $\lambda$) which regulates the long distance logarithmic divergence that commonly arises in such 2D Stokeslet flows. Beyond this length scale, the solvent stress dominates the in-plane membrane stress, regulating the logarithmic divergence. Membranes arising in nature typically have confined geometries with nonzero curvature. The role of membrane curvature and topology on the dynamics of embedded inclusions have been explored in many recent works \cite{henlev2008,henlev2010,wg2012, wg2013, atzbergershape, dan2016, atz2016, atz2018,atz2019, sosm20,atz22,sarthak22,smn22}. In the context of tubular membranes which is the focus of this paper, Daniels and Turner \cite{dt07} have shown that the particle mobility and diffusion coefficient feature a logarithmic dependence on tube radius R. A detailed and insightful analysis by Henle and Levine \cite{henlev2010} showed that the longitudinal
mobility of cylindrical membranes (in the limit of thin tubes) coincides with the logarithmic dependence on tube radius R predicted by Daniels and Turner. This has been experimentally demonstrated in a beautiful work by \cite{bss11}, where slow protein diffusion was observed in tubular membranes with smaller radii, consistent with the aforementioned theoretical predictions.\\\\
Membrane inclusions are self-driven motors that can perform mechanical work utilizing chemical energy. The stress generated by such force-free inclusions is essentially similar to that of force dipoles to leading order, as demonstrated by numerical simulations \cite{mik13}. A recent work \cite{mnk} showed aggregation of force dipoles in flat fluid membranes  via mutual hydrodynamic interactions, under suitable confinement of the external embedding fluids. These results have been subsequently extended to spherical fluid membranes \cite{sarthak22}. Pair dynamics of force dipoles in odd-viscous fluids have also been studied recently \cite{hk22}, see also Ref.~\cite{hk22rev} for a nice review of the field. In this paper, we construct an analytic solution of the fluid flow sourced by a point force dipole in a cylindrical membrane geometry of radius R (in the limit of infinitely long, thin cylinder). Besides being of experimental relevance \cite{bss11}, such  tubular membranes feature an anisotropic mobility tensor, in contrast to flat and spherical membranes \cite{henlev2010}. The purpose of the current study is to investigate the in-plane hydrodynamic interactions of a pair of force dipoles in the tubular fluid membrane surface, allowing momentum exchange with the external embedding fluids, please see Fig.~(\ref{3d}) for some visual illustrations of the dynamics. We formulate the dynamics of force dipoles on the cylinder following the formalism developed in \cite{sarthak22}. Focusing on pusher-type dipoles, we find that a mutually perpendicular dipole pair generically move together along helical geodesic trajectories on the tubular fluid membrane. The cylindrical geometry breaks the in-plane rotational symmetry of the membrane and creates significant differences in the flows along the compact angular and the long axial directions of the cylinder. For example, the flow along the compact angular direction of the cylinder is characterized by a \textit{local} rigid rotation term, which is independent of the angular coordinate but decays exponentially along the cylinder axis, with a decay length $\sqrt{\lambda R}$, see Eq.~(\ref{vstkhc}) where we present the Stokeslet flow. This term is absent in the flow along the non-compact axial direction of the cylinder. Due to this feature of the flow, we observe that a dipole pair initially interacting along the axial $\hat{z}$ direction exhibits an overall drift along the compact angular direction $\hat{\theta}$ of the cylinder surface. Our numerical simulations suggest that this drift increases \textit{linearly} with time for the dipole pair.\\
The effects of curvature can be understood as follows: for a generic curved, compact membrane in the setup described in this paper (see Eqs.~(\ref{systemeq}) in Sec.~\ref{mhd}), one expects two types of curvature contributions to the flows sourced by membrane bound inclusions. The first contribution is an explicit curvature contribution to the equations (see Eq.~(\ref{systemeq})) which can source the in-plane membrane flows. This is absent for a cylindrical membrane which has zero Gaussian curvature. The second contribution is extrinsic in nature and arises because the membrane geometry modifies the non-local interactions between any two points on the membrane surface via the external fluids. Thus, for a tubular fluid membrane, even in the absence of Gaussian curvature, the cylindrical geometry has considerable impact on the momentum exchange between the membrane and external fluids (please see Sec.~\ref{mhd} and Appendix \ref{aderiv} for detailed computation). What we observe in the simulations is a manifestation of these effects from extrinsic geometry and the compact topology of the cylinder in the angular $\hat{\theta}$ direction. It may be useful to comment on these results from the perspective of spherical and flat membranes. For example, for a Stokelet flow on the sphere, at high curvature there exists a soft mode that causes the entire membrane fluid, along with the internal fluid, to rotate like a rigid body and exhibit \textit{global} rotation \cite{henlev2010,sarthak22}. In contrast, the corresponding Stokeslet flow on an infinitely extended cylinder cannot support such a global motion, since the velocity field has to decay at infinity along the axis of the tube. However, for a force oriented along the angular direction, one  still has a \textit{local} rigid rotation, independent of the cylinder angular co-ordinate $\theta$ but decaying along the z-axis of the cylinder. Note that for the case of Stokeslet flow in infinitely extended flat membranes, neither global nor local rigid rotation of the membrane fluid is possible.\\\\
The paper is organized as follows: In Sec.~\ref{mhd} we summarize the membrane hydrodynamics appropriate for
curved geometries. In the limit of thin tubular membranes, we provide analytic solution for the flow sourced by a point force (Stokeslet), followed by the flow sourced by a point force dipole. The general solution for a cylindrical membrane of arbitrary radius (along with a detailed derivation) is presented in Appendix \ref{aderiv}.  In Sec.~\ref{dynsim}, we formulate the dynamical equations relevant for the regime where hydrodynamic interactions dominate over thermal fluctuations and perform numerical simulations of pair dynamics of force dipoles. We identify the self-propelling helical geodesics for mutually perpendicular dipole pair. Next, we define and identify the ``drift" arising from the \textit{local} rigid rotation. We end with a summary and future outlook in Sec.~\ref{cncl}. The main text is supplemented by Appendix \ref{aderiv} where the equations of membrane hydrodynamics are solved for a cylindrical geometry to arrive at Eq.~(\ref{vstkhc}) of main text. For completeness, Appendix \ref{amap} presents a more exhaustive set of trajectories for the two-dipole system found by varying two key parameters in the model, namely the initial relative separation and orientation.
\section{In-plane Membrane Hydrodynamics}
\label{mhd}
The setup we consider in this paper is based on the pioneering works \cite{saff1,saff2}, generalized to curved geometries \cite{henlev2008,henlev2010}. We consider a two dimensional, curved, incompressible fluid membrane of viscosity $\eta_{2d}$  surrounded by external embedding fluids, both outside and inside the membrane surface. We also restrict to situations where the membrane geometry is fixed and consider only tangential flows in the membrane. The membrane fluid is also assumed to be impermeable to external fluids. However, the membrane can exchange momentum with the external fluids, hence it is quasi-2D. Such a fluid model is described by the following set of equations:
\beqa
&D^\alpha v_\alpha =0,\nn\\ 
&\sigma^{ext}_{\alpha} = \underbrace{-\eta_{2D} \left( K(\vec{x})~ v_\alpha  + D^\mu D_\mu v_\alpha \right) + D_\alpha p}_{Membrane~stress}~~
+ \underbrace{T_\alpha}_{External~stress}, \nn\\[6pt]
&~ \nabla \cdot {\bf v}_{\pm} = 0, ~~~\eta _{\pm}\nabla^2 {\bf v}_{\pm} = \nabla_{\pm} p^{\pm},\nn\\
& T_\alpha= \sigma_{\alpha r}^{-}|_ {r=R} -\sigma_{\alpha r}^+ |_ {r=R}, ~~ \sigma_{ij}^\pm =\eta_\pm \left(D_i v_j^\pm +D_j v_i^\pm\right)- g_{ij} p_{\pm},\nn\\[6pt]
&{\bf v}_\pm | _{r=R}=v.
\label{systemeq}
\eeqa
We now describe the notation used in the above set of equations. First, we reserve Greek indices for co-ordinates of the 2D membrane surface while Latin indices are used for 3D co-ordinates to describe velocity fields of the external 3D embedding fluids. The first equation in Eqs.~(\ref{systemeq}) is the incompressibility equation for the 2D membrane fluid velocity field $v_\alpha$ and  $D$ is the covariant derivative compatible with the membrane surface metric $g_{\alpha \beta}$. The second equation in Eqs.~(\ref{systemeq}) is the equation of stress balance at the membrane surface. Here $\sigma^{ext}_{\alpha} $ is the local stress exerted by the motors (inclusions) embedded in the membrane and this is balanced by the membrane stress and the stress from the external embedding fluids (denoted by $T_\alpha$) as indicated in the equation. The co-ordinate $x$ denotes generic membrane surface co-ordinates.  We work in the regime of low Reynolds hydrodynamics where the 2D membrane fluid flows are dictated by Stokes equations, with viscosity $\eta_{2d}$. $K(\vec{x})$ is the local Gaussian curvature for the given membrane geometry and $p$ is the 2D membrane pressure.  The membrane fluid is surrounded by external 3D embedding fluids with viscosities $\eta_+$ ( $\eta_-$) and pressure  $p_+$ ($p_-$) for the outer (inner) fluid  respectively. The third equation is the Stokes flow equations for these external incompressible fluids in 3D. The fourth equation gives the expression for the traction vector $T_\alpha$ that appeared in the stress balance equation.  It is constructed from the fluid stress tensor of the external fluids $\sigma^{\pm}$ with $g_{ij}$ being the ambient 3D metric. The co-ordinate $r$ represents the direction normal to the membrane surface.  Finally the last equation is the no-slip boundary condition which ensures that the 2D membrane fluid flow matches the external fluid velocity at the membrane surface described by $r=R$. We also define two length scales
\beqa
\lambda_+ =\frac{ \eta_{2d}}{\eta_+},~~~\lambda_- =\frac{ \eta_{2d}}{\eta_-}.\nn
\eeqa
The system of equations Eqs.~(\ref{systemeq}) can be solved analytically for surfaces of constant Gaussian curvature. For surfaces of varying Gaussian curvature, one requires a numerical treatment similar to \cite{atz2018,sosm20}. In both situations, the strategy to solve the system is to decompose the in-plane membrane velocity field in terms of eigenmodes of the Laplace Beltrami operator of the surface under consideration. One can then solve for the coefficients of these eigenmodes in terms of the point source $\sigma^{ext}_{\alpha}$ and finally performing an inverse transform to write the relevant Green's function in real space. We refer to \cite{henlev2008,henlev2010} for the detailed steps to solve Eq.~(\ref{systemeq}), as well as Ref.~\cite{sarthak22}. In this paper, our focus is on infinitely extended cylindrical membranes where the Gaussian curvature vanishes, facilitating analytic treatment. For simplicity, we will assume that the external embedding fluids inside and outside the cylindrical membrane have the same viscosity $\eta_+=\eta_- \equiv \eta$ which gives rise to a unique Saffman length $\lambda$ 
\beqa
\lambda= \frac{\eta_{2d}}{\eta}.
\eeqa
In this notation, the high curvature regime \footnote{ We denote the geometry in this limit as "tubular membrane" while the generic situation with arbitrary curvature as "cylindrical membrane".} arises when $\frac{\lambda}{R} \gg 1$.   In this limit, we find new analytic expressions for the Stokeslet flow in the tubular membrane of radius R coupled to external embedding fluids (see Appendix \ref{aderiv} for a detailed computation). For a point force of unit strength $F_0=1$ situated at the origin and making an angle $\alpha_0$ with respect to $\hat{\theta}$ (ie. $\vec{F}_{0} \cdot \hat{\theta}_{0}=\cos \alpha_0$ and $ \vec{F}_{0} \cdot \hat{z}=\sin \alpha_0$), the Stokeslet flow $\bm{v}^{Stk}$ in the limit of high curvature (tubular membrane) is given by

\begin{align}
&v_\theta^{Stk}[\theta,z]=\frac{1}{8 \pi \eta_{2d} R}  \left(-\frac{z \sin \alpha_0 \sin \theta}{ \cos \theta- \cosh \frac{z}{R}}+\cos \alpha_0 \left(-|z| \frac{e^{-\frac{|z|}{R}} -\cos \theta}{ \cos \theta - \cosh \frac{|z|}{R}}-R~ \log \left[1-2 e^{-\frac{|z|}{R}} \cos \theta+e^{-\frac{2 |z|}{R}}\right]+ \sqrt{2 R \lambda}~ e^{-\frac{\sqrt{2} |z|}{\sqrt{\lambda R}}}\right)\right)\nn\\
& v_z^{Stk}[\theta,z]=\frac{1}{8 \pi  \eta_{2d} R} \left(-\frac{z \cos \alpha_0 \sin \theta}{\cos \theta-\cosh \frac{z}{R}}+\sin \alpha_0 \left(|z| \frac{e^{-\frac{|z|}{R}} -\cos \theta}{ \cos \theta - \cosh \frac{|z|}{R}}-R~ \log \left[1-2 e^{-\frac{|z|}{R}} \cos \theta+e^{-\frac{2 |z|}{R}}\right]\right)\right)
\label{vstkhc}
\end{align}
The last term appearing in $v_\theta^{Stk}$  arises from the zero mode along the compact angular direction of the tube and is independent of $\theta$ but decays along the axis of the cylinder (z direction) with a length scale $\sqrt{\lambda R}$. For a point force directed along $\hat{\theta}$ ie. $\alpha_0 =0$, this creates  a \textit{local}  rigid rotation, as opposed to a \textit{global} rotation seen in the spherical fluid membrane \cite{henlev2010,sarthak22}. Note that such a zero mode contribution is absent when the force acts along the long z-axis of the tube ie. $\alpha_0=\pi/2$ where the decay length for the flow is controlled by the radius R of the tube.
% This also leads to anisotropic  longitudinal ($\mu_z$) and transverse ($\mu_\theta$) mobilities \cite{henlev2010}:
%\begin{equation}
%\begin{gathered}
%\lim _{R / \lambda \rightarrow 0} \mu_z=\frac{1}{4 \pi \eta_{2d}}\left[\ln \left(\frac{R}{a}\right)+\frac{1}{2}\right], \\
%\lim _{R / \lambda \rightarrow 0} \mu_{\theta}=\frac{1}{4 \pi \eta_{2d}}\left[\sqrt{\frac{\lambda}{2 R}}+\ln \left(\frac{R}{a}\right)-\frac{1}{2}\right] 
%\label{mob}
%\end{gathered}
%\end{equation}
%where a is the inclusion size. 
The formalism to construct the corresponding flow sourced by a point force dipole in curved membranes has been discussed in \cite{sarthak22}. In particular, since the cylindrical geometry is intrinsically flat, the analysis simplifies, allowing us to construct the force dipole flow with relative ease. The flow at the location $(\theta,z)$ on the tubular membrane sourced by a force dipole of strength $\kappa$, situated at $(\theta_0,z_0)$ and making an angle $\alpha_0$ with respect to $\hat{\theta}$ is given by
\beqa
v_\theta^{dipole}[\theta,z]= \kappa \left(\frac{\cos \alpha_0}{R} ~ \partial_{\theta_0} + \sin \alpha_0~ \partial_{z_0}\right) v_\theta^{Stk}[\theta-\theta_0,z-z_0] \nn\\
v_z^{dipole}[\theta,z]= \kappa  \left(\frac{\cos \alpha_0}{R} ~ \partial_{\theta_0} + \sin \alpha_0~ \partial_{z_0}\right) v_z^{Stk}[\theta-\theta_0,z-z_0]
\label{vdp}
\eeqa

where $ v_\theta^{Stk}$ and $ v_z^{Stk}$ are defined in Eq.~(\ref{vstkhc}). For a spherical membrane, the oppositely oriented forces in the dipole leads to a cancellation of the aforementioned  global rotation. However, in a cylindrical membrane, the in-plane rotational symmetry is broken and a finite \textit{local} rigid rotation term survives (Mathematically, the point force dipole solution arises by taking a directional derivative of the Stokeslet flow. For a spherical membrane, it was shown in \cite{sarthak22} that the directional derivative kills the  zero mode. However in the cylindrical membrane, this mode is spatially dependent and hence is non-zero in the force dipole solution).
\begin{figure}[h]
\includegraphics[width=6.5cm]{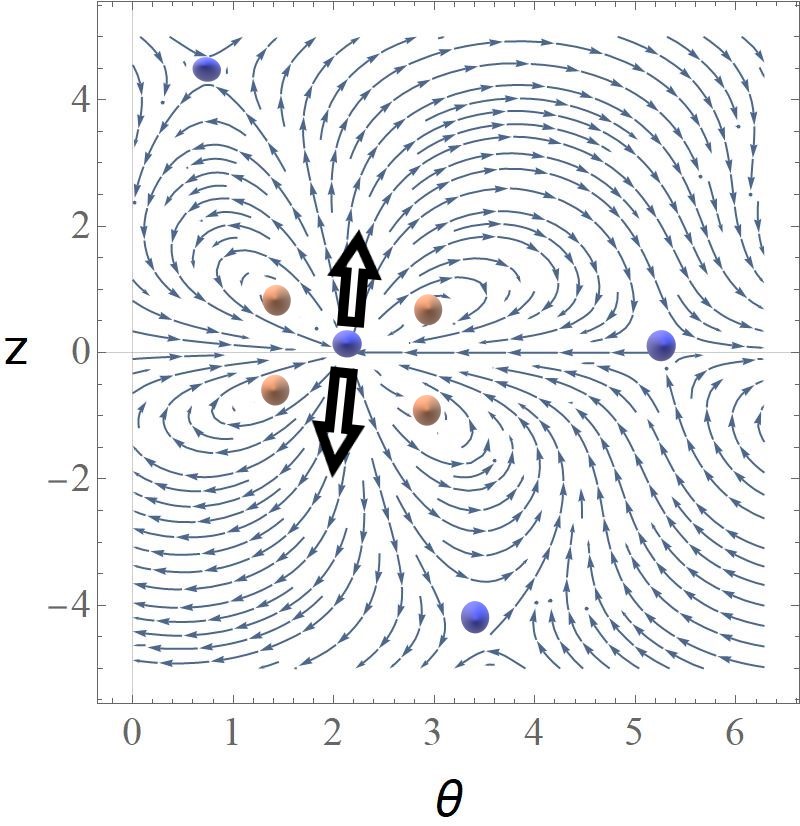}
\caption{ Plot of the flow Eq.~(\ref{vdp}) created by a force dipole (marked by black arrows), where we have mapped the cylinder $(\theta,z)$ coordinates to the plane. The blue dots represent the core of saddles and the brown dots represent vortex cores.}
     \label{dpflow}
\end{figure}
In the context of pair dipole interactions in the tubular fluid membrane, one can easily observe non-trivial dynamics caused by the  \textit{local} rigid rotation. If we start with an initial configuration of two dipoles separated along the z-axis of the cylinder by a distance $L_0$, we expect the interacting dipoles to drift along the angular $\hat{\theta}$ direction of the tube due to the local rigid rotation. This drift is absent in spherical and flat membranes. Our numerical investigations suggest that the dipole pair drifts \textit{linearly} with time.\\\\
\textbf{Comments on topological aspects of streamlines:} We end this section with some comments on the topological aspects of the streamlines created by the point force dipole in the tubular fluid membrane. We observe a quadrupolar flow surrounding the dipole location, which adds to a net index of +4 (see Fig.~(\ref{dpflow}), where the vortex cores are marked by brown dots). At the core of the dipole, we have a saddle and the neighboring flow field features three additional saddles, marked by blue dots. Thus the net topological index associated to the flow field is zero, which is consistent with the Euler characteristic of the cylinder (Poincar\'e index theorem). In contrast, the flowfield of a dipole in a spherical membrane features only one saddle, such that the net index is two, which is the Euler characteristic of a sphere, as shown in \cite{sarthak22}. The corresponding flow in flat membrane features no additional saddles \cite{mnk}.

\section{Dynamical Equations and Simulation Results}
\label{dynsim}
\begin{figure}[htbp!]

\begin{subfigure}{0.49\textwidth}
\includegraphics[width=0.8\linewidth]{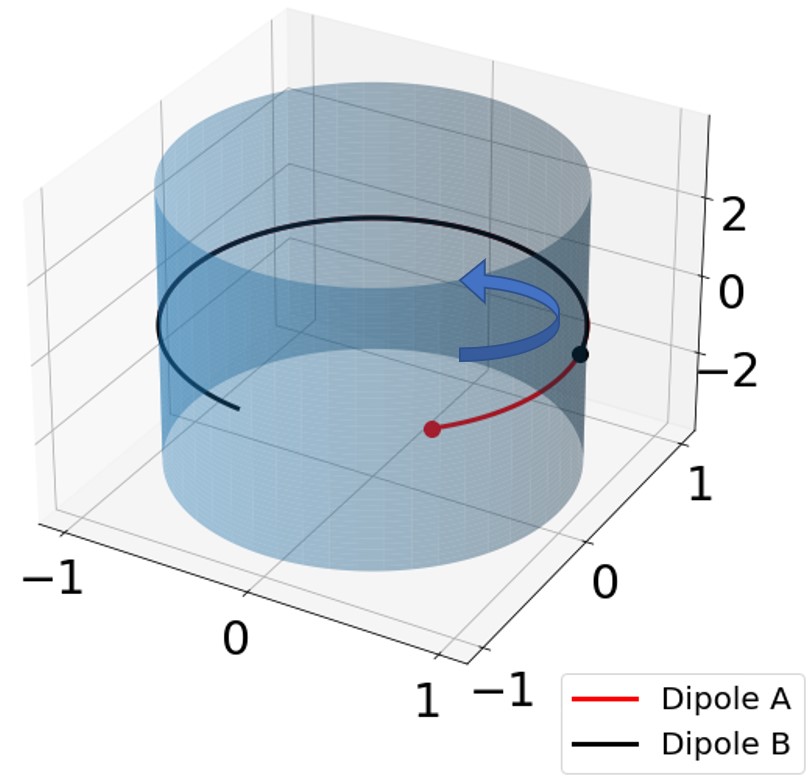}
  
  \caption{Case A: $r_1 = (-\frac{\pi}{4},0,0)$, $r_2 = (-\frac{\pi}{4}+1, 0, \frac{\pi}{2})$.}
  \label{3d_a}
\end{subfigure} 
\begin{subfigure}{0.49\textwidth}
  \includegraphics[width=0.8\linewidth]{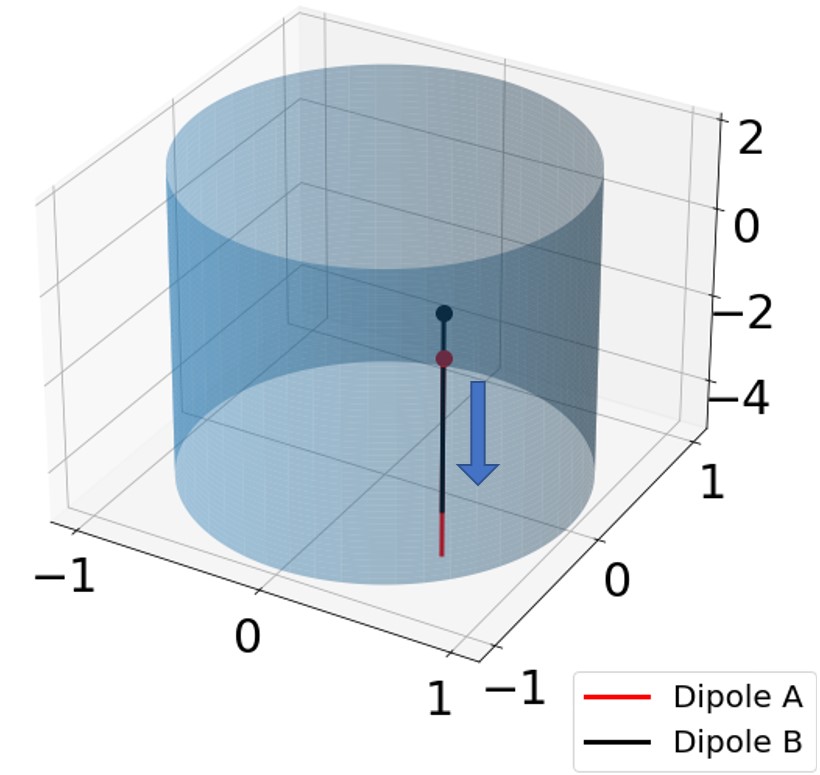}
  
  \caption{Case B: $r_1 = (-\frac{\pi}{4},0,0)$, $r_2 = (-\frac{\pi}{4}, 4, \frac{\pi}{2})$.}
  \label{3d_b}
\end{subfigure}
\begin{subfigure}{0.49\textwidth}
\includegraphics[width=0.8\linewidth]{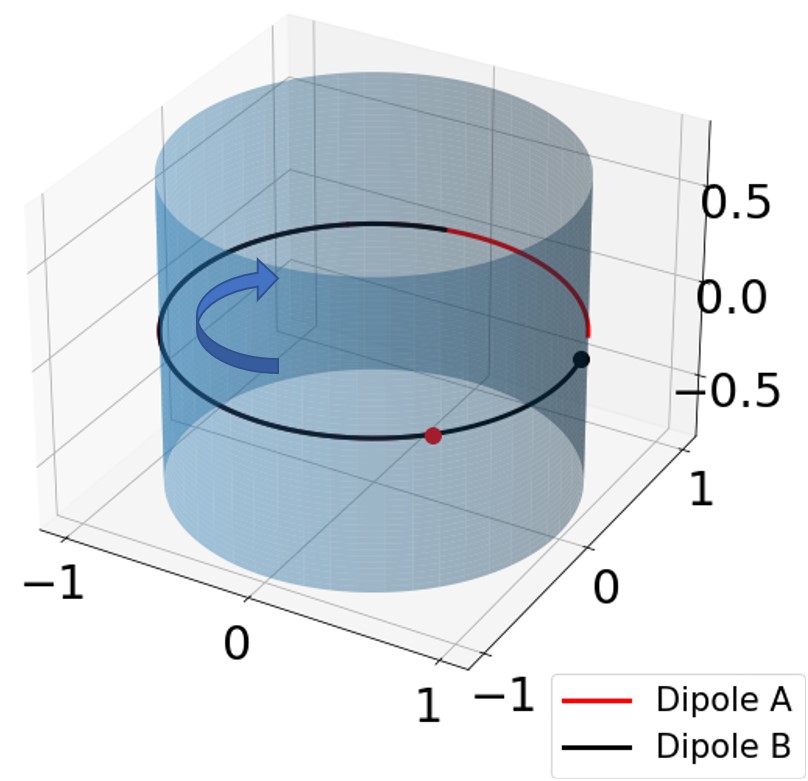}
  
  \caption{Case C: $r_1 = (0,0,\frac{\pi}{4})$, $r_2 = (1, 0, 0)$.}
  \label{3d_c}
\end{subfigure} 
\begin{subfigure}{0.49\textwidth}
  \includegraphics[width=0.8\linewidth]{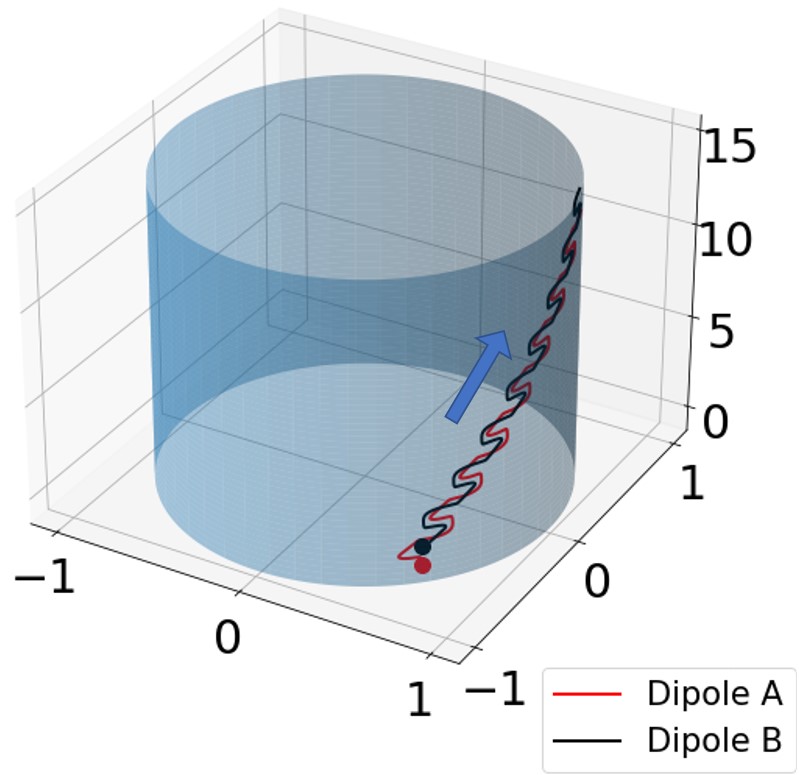}
  
  \caption{Case D: $r_1 = (0,0,\frac{\pi}{4})$, $r_2 = (0, 1, 0)$.}
  \label{3d_d}
\end{subfigure}
\begin{subfigure}{0.49\textwidth}
\includegraphics[width=0.8\linewidth]{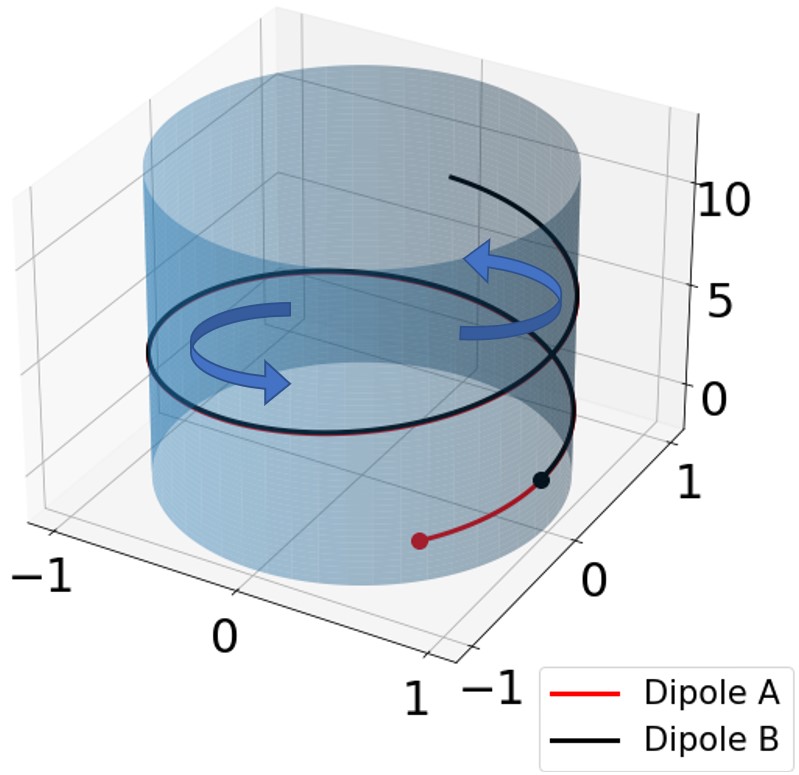}
  
  \caption{Case E: $r_1 = (0,0,\frac{\pi}{4})$, $r2 = (\frac{1}{\sqrt{2}}, \frac{1}{\sqrt{2}}, \frac{3\pi}{4})$.}
  \label{3d_e}
\end{subfigure} 
\begin{subfigure}{0.49\textwidth}
\includegraphics[width=0.8\linewidth]{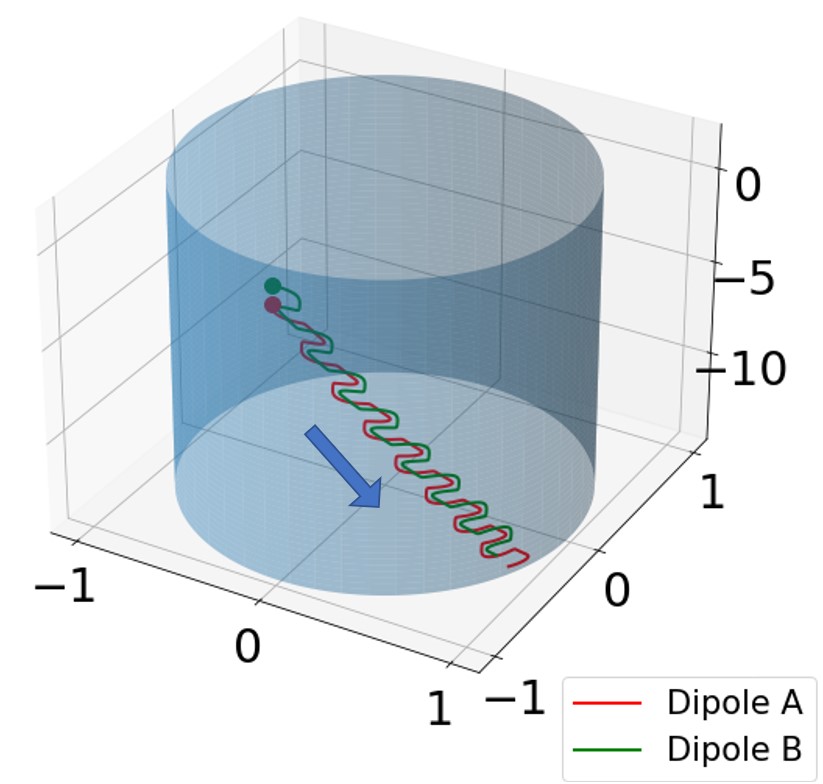}
  
  \caption{Case F: Same as Case D, except $\kappa = -1$ for the second dipole (green color). }
  \label{3d_f}
\end{subfigure}
\caption{3D plots of trajectories of dipole pair for six cases A-F. The colored dots indicate the initial position of the dipoles, giving rise to the dipole trajectories marked with same color as the dipole. Arrows indicate the direction of motion of the dipoles. For case F, green colored  dipole has strength $\kappa=-1$. For all simulations, $\frac{\lambda}{R} =100$.} 
\label{3d}
\end{figure}
In this section, we formulate the equations governing the dynamics of a system of point force dipoles embedded in the tubular fluid membrane. Each dipole translates according to the local flow induced by the superposition of flows sourced by the remaining dipoles (via linearity of Stokes equations). The dipole also rotates at a rate $\dot{\alpha}$ according to the local vorticity of the flow sourced by the rest of the dipoles. In this framework, we focus on the hydrodynamic interactions of a pair of such dipoles and highlight the interesting aspects arising from the cylindrical geometry of the membrane. We also explore some of these aspects in the context of multi-dipole systems. The motion of the $i^{\textit{th}}$ dipole with co-ordinate $(\theta_i,z_i)$ is thus given by:
\beqa
R ~ \dot{\theta}_i=  \sum_{j \neq i}^N  \bm{v}^{dipole}_{\theta}, 
\hspace{1cm} \dot{z}_i=  \sum_{j \neq i}^N   \bm{v}^{dipole}_{z}, \hspace{1cm}\dot{\alpha}_i=  \sum_{j \neq i}^N \underbrace{\frac{1}{2} \left(\nabla^{cyl}_i \times \bm{v}^{dipole}\right).\hat{r}_i}_{\text{Vorticity induced rotation}} 
\label{dynmeq}
\eeqa

\begin{figure}[H]
\hspace{-2cm}
\begin{subfigure}{0.65\textwidth}
\includegraphics[width=\linewidth]{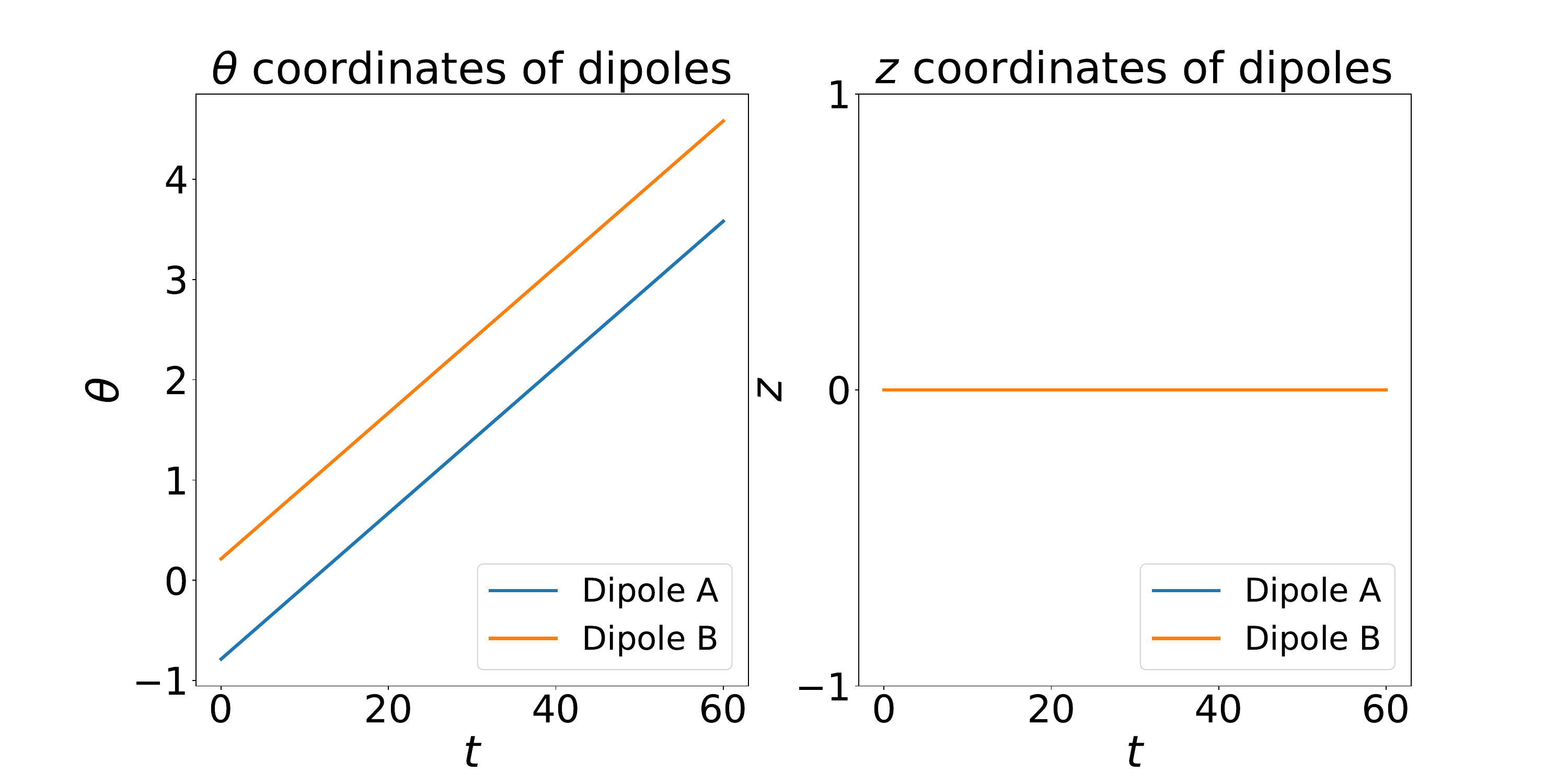}
  
  \subcaption{Case A}

\end{subfigure} \hspace{-1.7cm}
\begin{subfigure}{0.65\textwidth}
  \includegraphics[width=\linewidth]{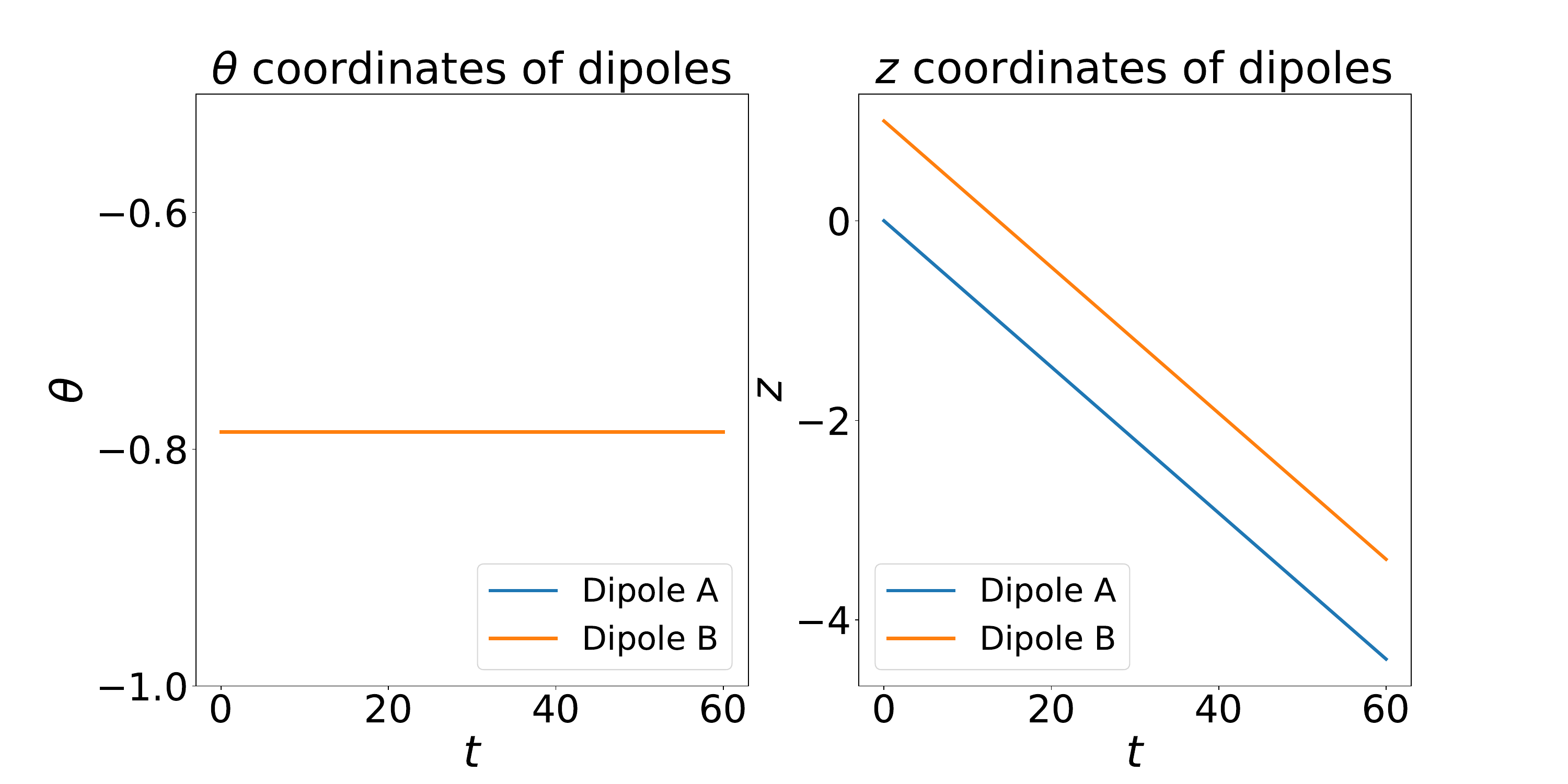}
  
  \subcaption{Case B}

\end{subfigure}
\end{figure}
\begin{figure}[H]\ContinuedFloat
 \hspace{-2cm}
\begin{subfigure}{0.65\textwidth}
\includegraphics[width=\linewidth]{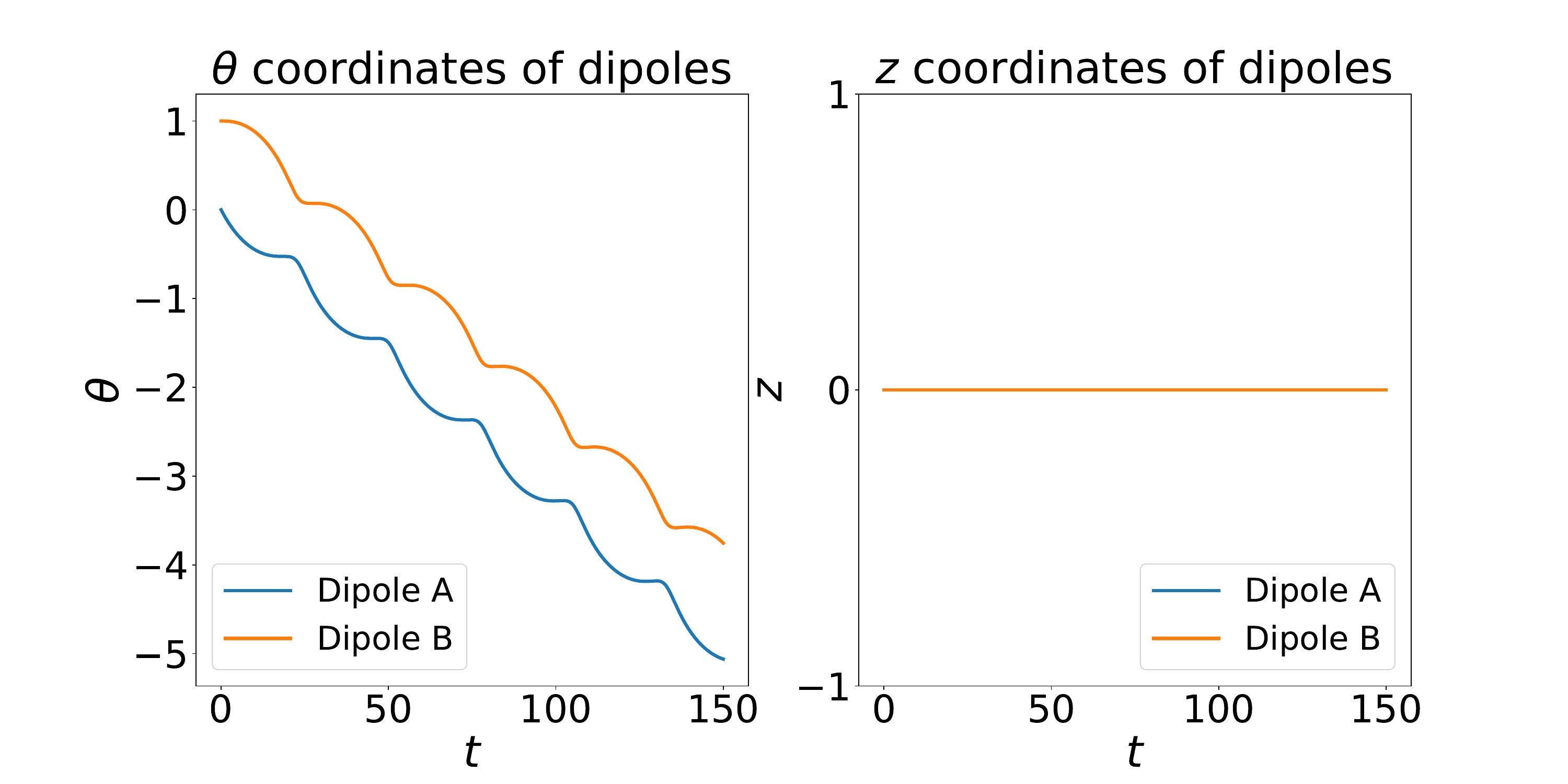}
  
  \subcaption{Case C}

\end{subfigure} \hspace{-1.7cm}
\begin{subfigure}{0.65\textwidth}
  \includegraphics[width=\linewidth]{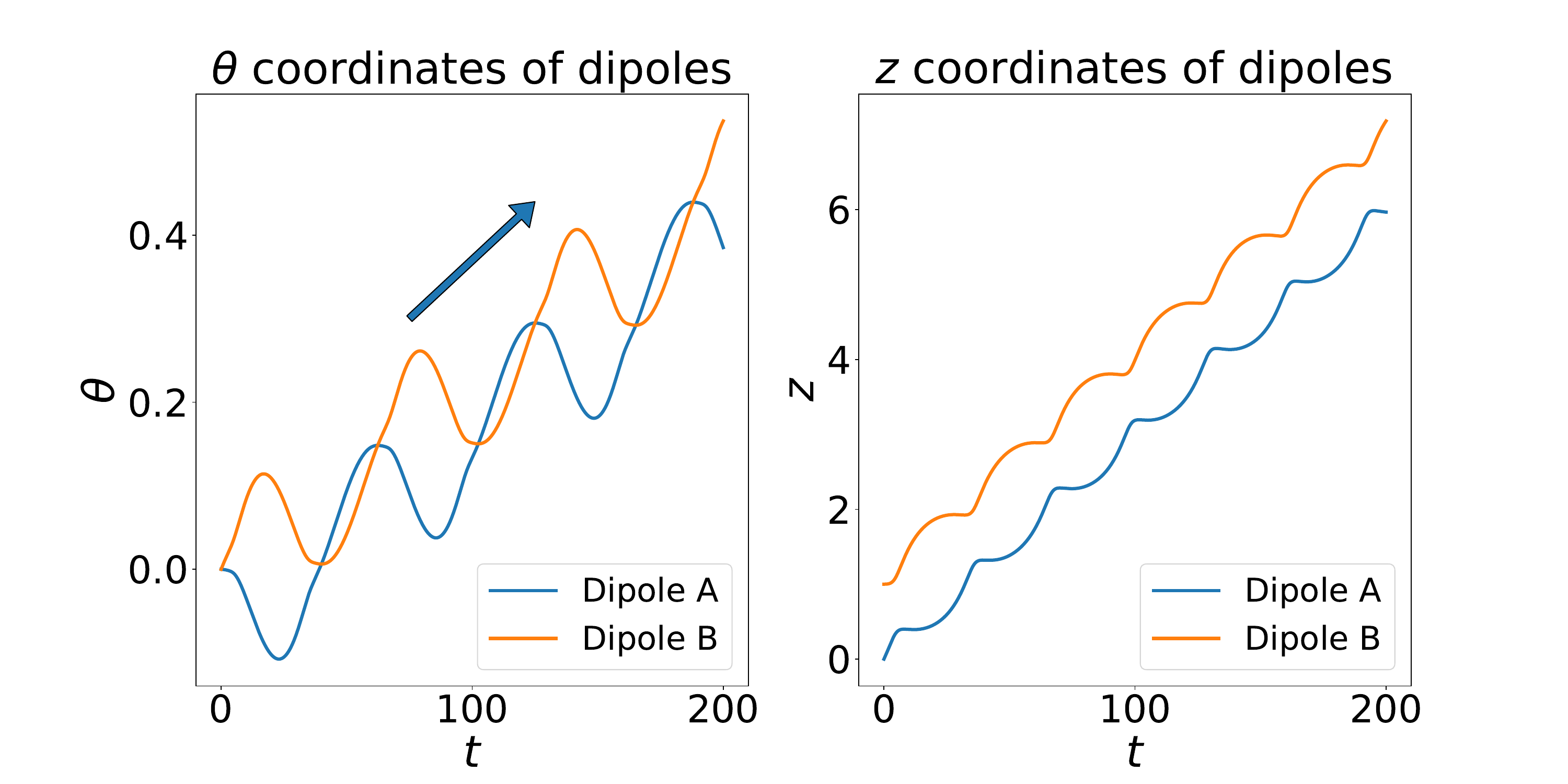}
  
  \subcaption{Case D}

\end{subfigure}
\end{figure}
\hspace{-2cm}
\begin{figure}[H]\ContinuedFloat
\hspace{-2cm}
  \begin{subfigure}{0.65\textwidth}
  \includegraphics[width=\linewidth]{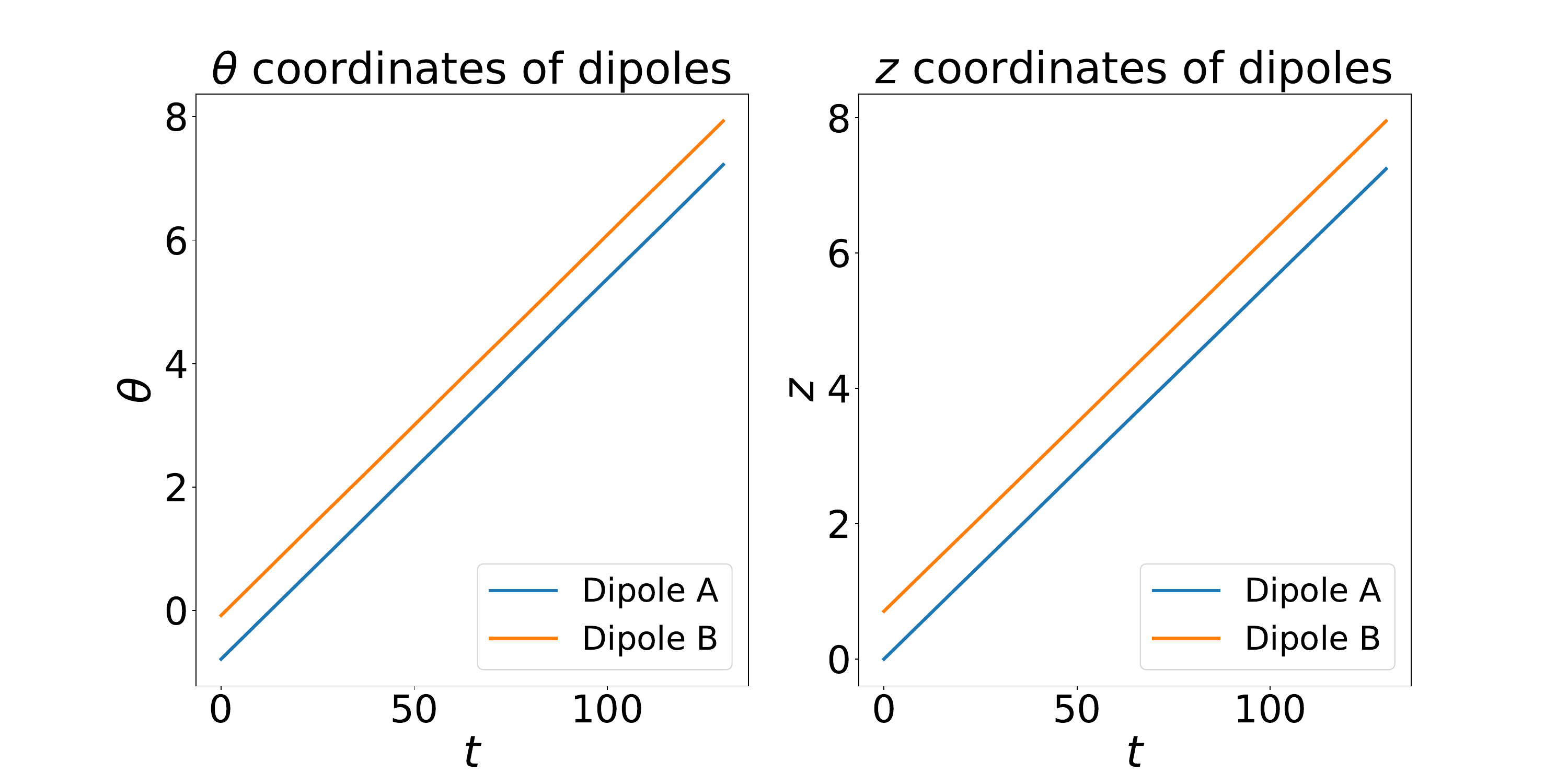}
  \subcaption{Case E}

\end{subfigure} 
\hspace{-1.7cm}
\begin{subfigure}{0.65\textwidth}
  \includegraphics[width=\linewidth]{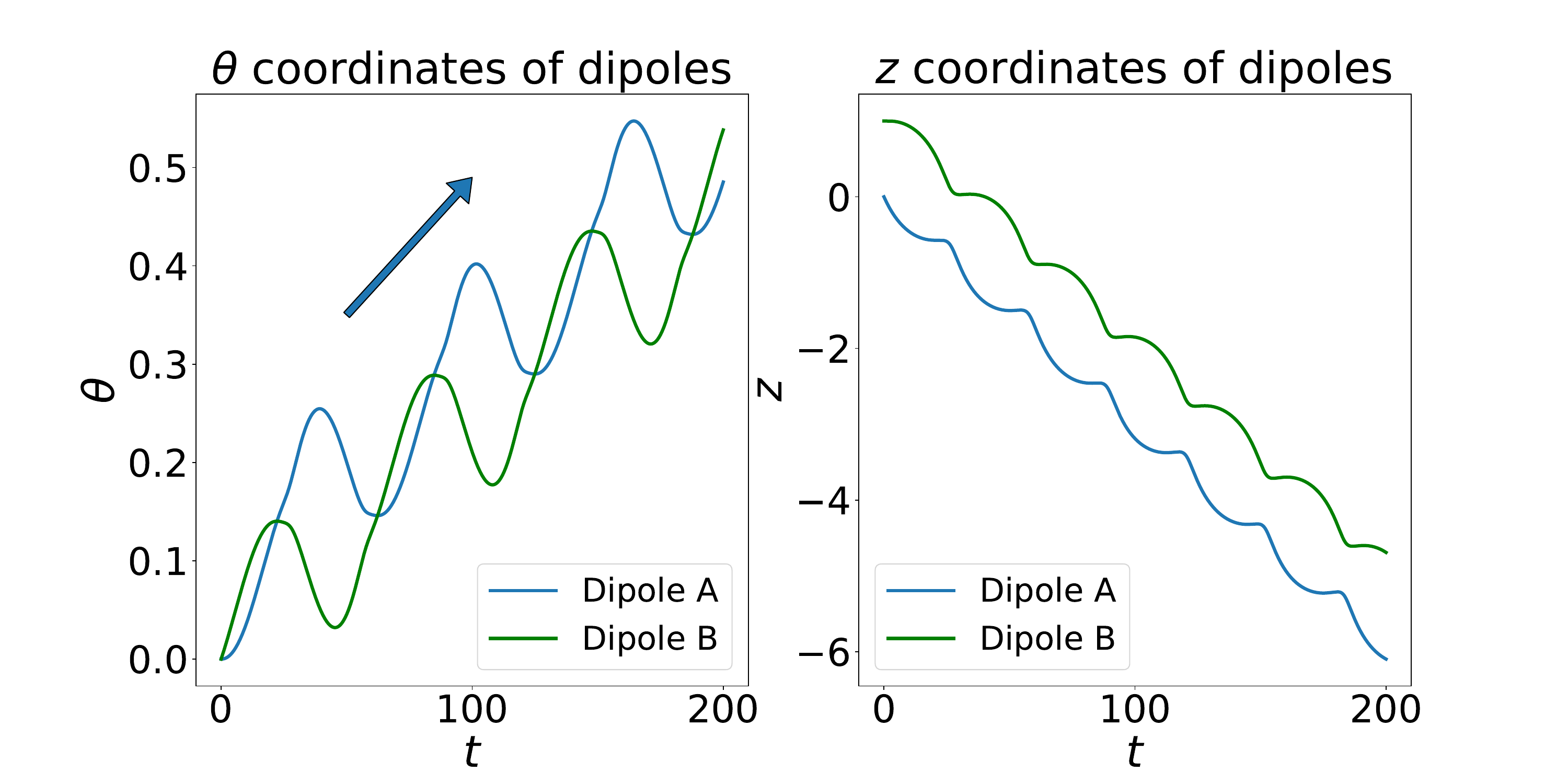}
  
  \subcaption{Case F}
  
\end{subfigure}
\caption{\label{zoomedfig} Zoomed figures of simulations of Cases A-F showing the temporal evolution of $\theta$ and $z$ coordinates of the dipole pair. The initial conditions and parameters are the same as those for Fig.~(\ref{3d}). For case F, the green color represents a dipole with strength $\kappa=-1$. The temporal evolution of orientation of dipole pair and the corresponding trajectories in $\theta-z$ plane are presented in Fig.~(\ref{azoomedfig}) of Appendix \ref{amap}. }

\end{figure}

\begin{figure}[h]
\centering
    \subfloat[\centering]{{\includegraphics[width=6.65cm]
  {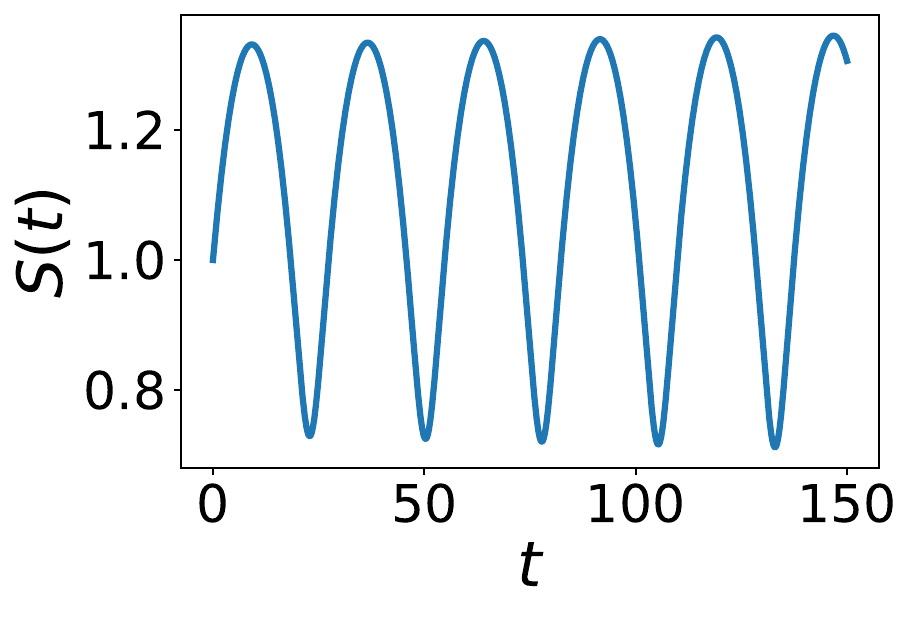} }}%
    \subfloat[\centering]{{\includegraphics[width=6.65cm]{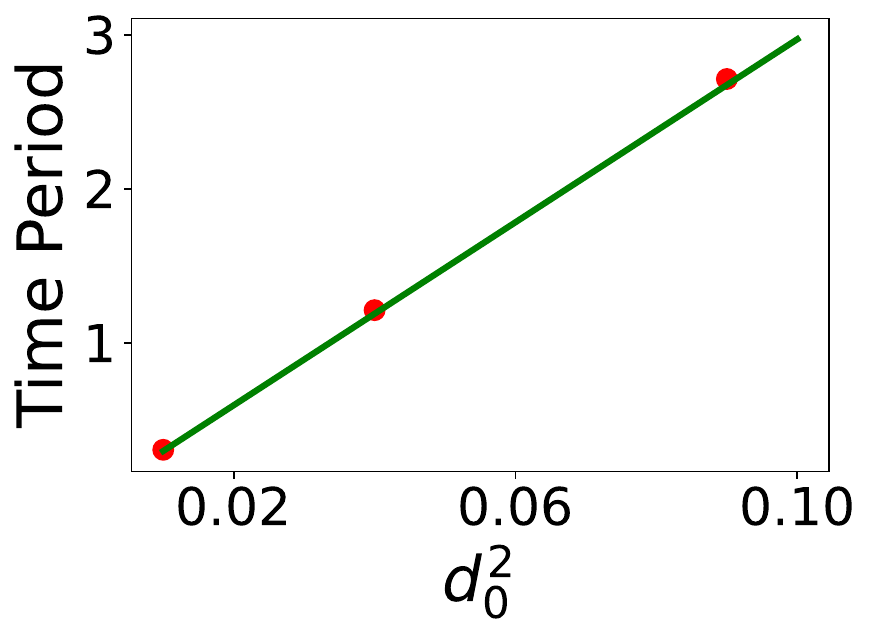} }}%
    \caption{a) Oscillations in the inter-dipole separation in Case C. The initial distance was set as 0.1.
    b) Linear fit showing the time period of the oscillations varying as $d_0^2$, where $d_0 = 0.1, 0.2, 0.3$ is the initial separation between the dipole pair.}
    \label{to_do}
\end{figure}

\begin{figure}[h]%
    \centering
    \subfloat[\centering]{{\includegraphics[width=6.65cm]
  {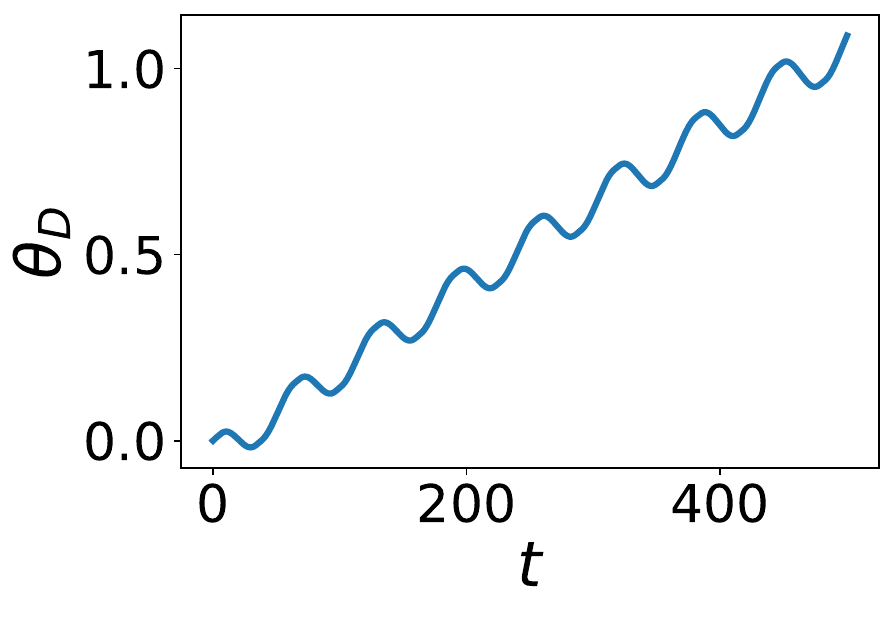} }}%
    \qquad
    \subfloat[\centering]{{\includegraphics[width=6.65cm]{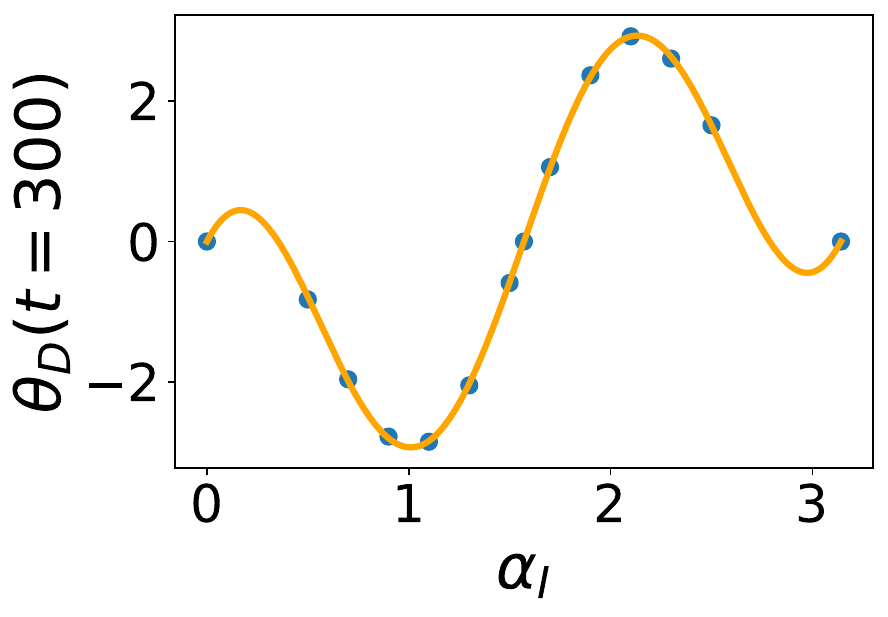} }}%
   % \subfloat[\centering]{{\includegraphics[width=5.3cm]{images/drift_r.png%} }}%
    \caption{a) Illustration of the drift $\theta_D$ in Case D (see main text)
    b) Variation of drift in $\theta_1$ at $t = 300$ against $\alpha_I$. The best fit curve is $(-4.1263 
 + 2.7852~(\alpha_I - \frac{\pi}{2})^2)\sin(2\alpha_I)$.}%
    
 \label{drift_fig}%
\end{figure}

\begin{figure}[htbp!]%
    \centering
    \subfloat[\centering]{{\includegraphics[width=5.65cm]
  {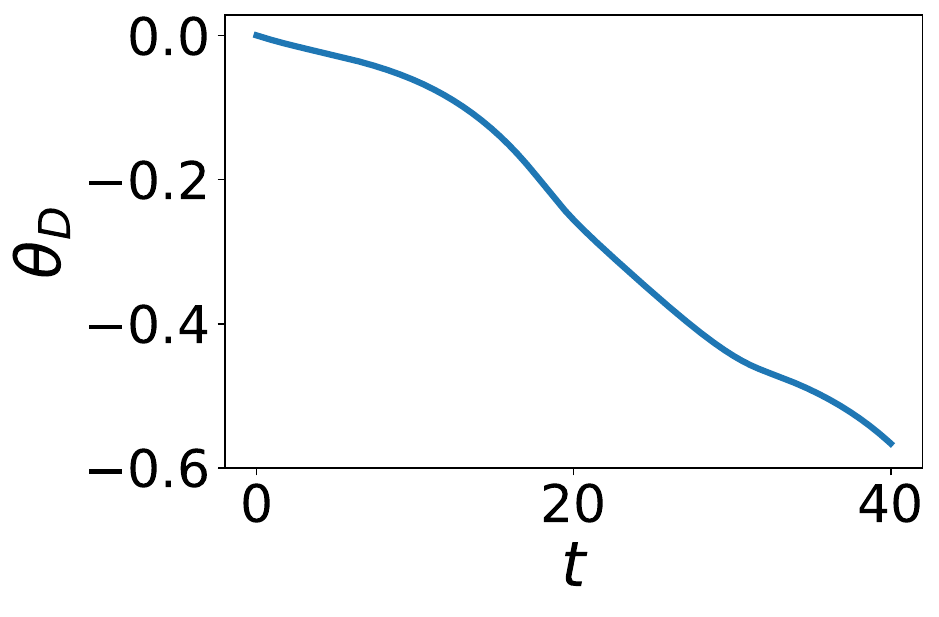} }}%
    \qquad
    \subfloat[\centering]{{\includegraphics[width=5.65cm]{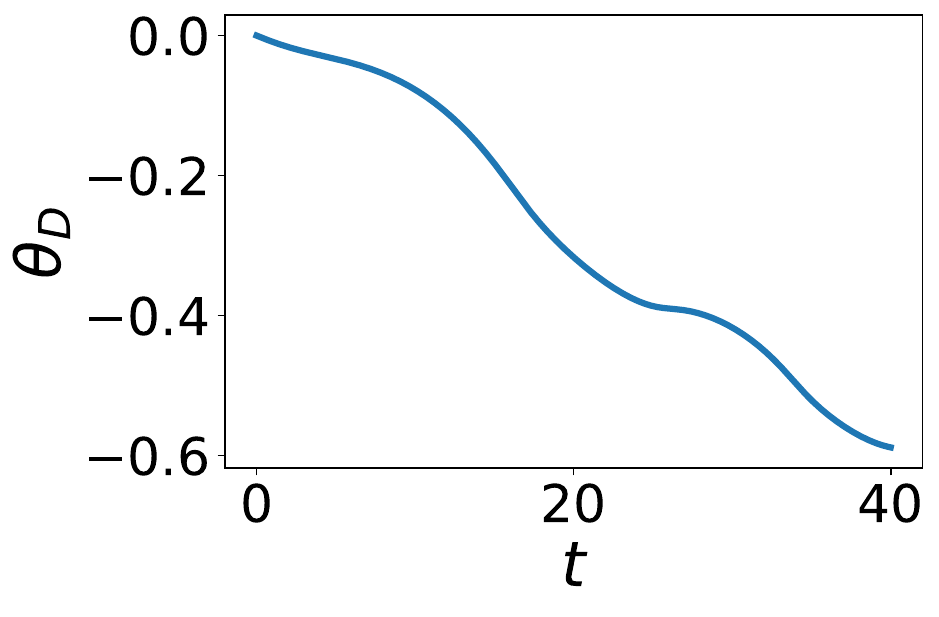} }}%
   % \subfloat[\centering]{{\includegraphics[width=5.3cm]{images/drift_r.png%} }}%
    \caption{a) Illustration of the drift $\theta_D$ for multi-dipole configurations of a) $r_1 = (0,0,0), r_2 = (0, 1, \frac{\pi}{4}), r_3 = (0,2,\frac{\pi}{4})$ and b) $r_1 = (0,0,0), r_2 = (0, 1, \frac{\pi}{4}), r_3 = (0,2,\frac{\pi}{4}), r_4 = (0,3,\frac{\pi}{4})$.}%
    \label{theta_d multi dipole}
\end{figure}

\begin{figure}[htbp!]
  \centering
  \includegraphics[width=\linewidth]{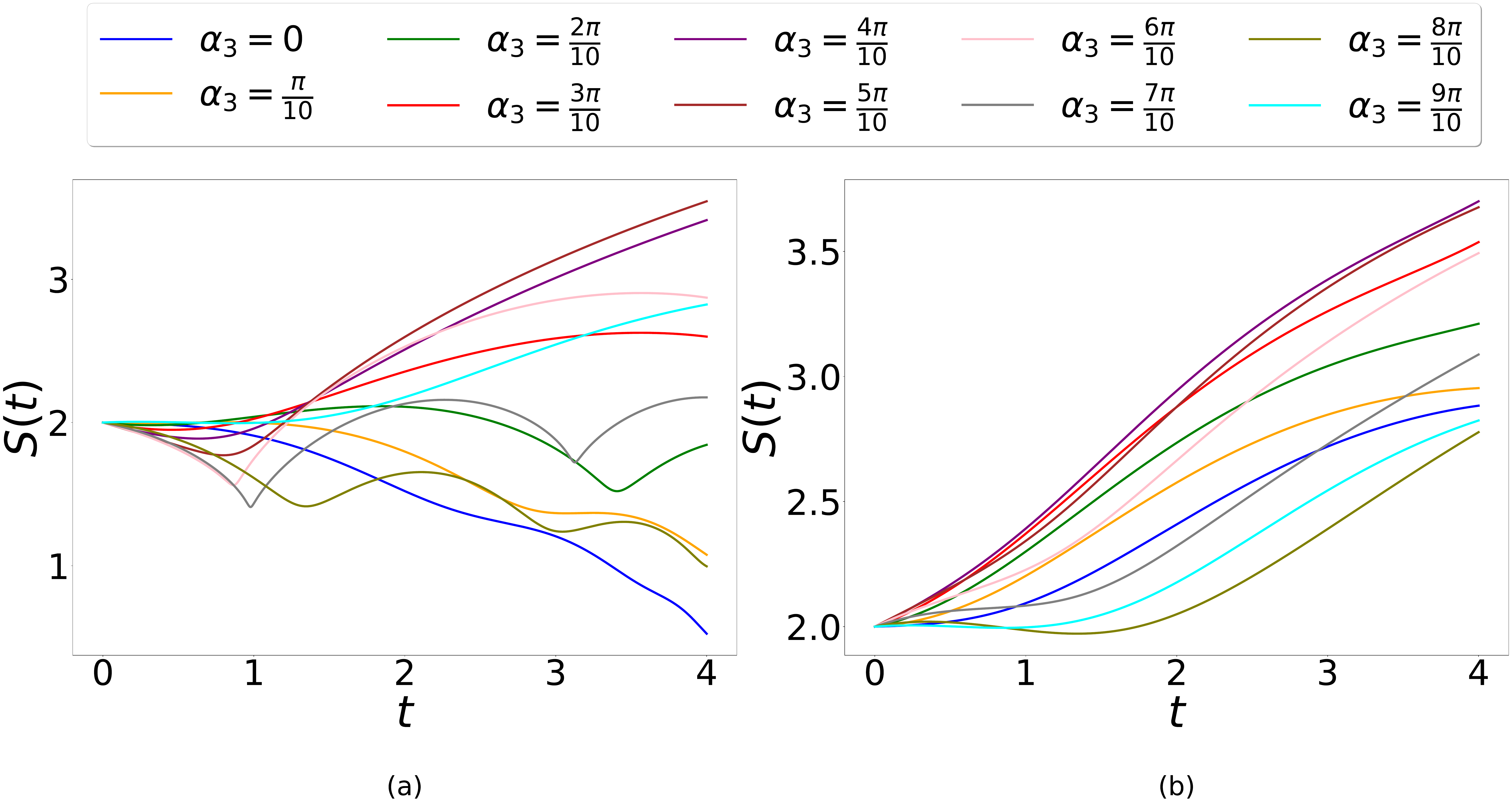}
  
  \caption{Sum of inter dipole distances for three-dipole configurations, initially situated along the a) Transverse, i.e., $r_1 = (0, 0, 0)$, $r_2 = (0.5, 0, \frac{\pi}{3})$, $r_3 = (1, 0, \alpha_3)$ and b) Longitudinal, i.e., $r_1 = (0, 0, 0)$, $r_2 = (0, 0.5, \frac{\pi}{3})$, $r_3 = (0, 1, \alpha_3)$ of the tubular membrane.. $\alpha_3$ is varied from 0 to $\frac{9\pi}{10}$. The plots for $\alpha_3 = a$ and  $\alpha_3 = a + \pi$ were found to be identical for all $a$.}
  \label{3dpdis}
\end{figure}

\begin{figure}[htbp!]
  \centering
  \includegraphics[width=\linewidth]{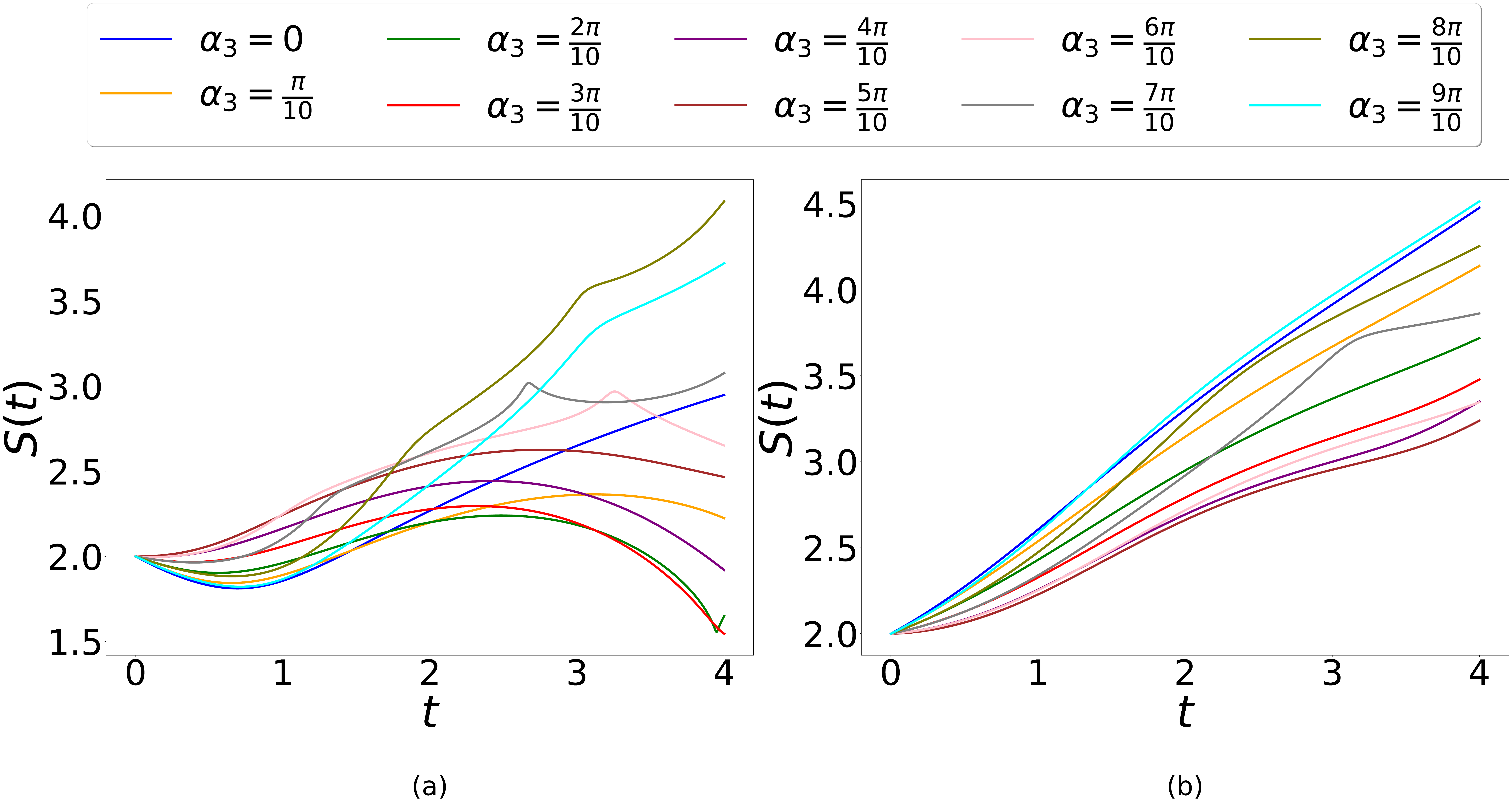}
  
  \caption{Sum of inter dipole distances for four-dipole configurations, initially situated along the a) Transverse, i.e., $r_1 = (0, 0, 0)$, $r_2 = (0.5, 0, \frac{\pi}{3})$, $r_3 = (1, 0, \frac{\pi}{3})$, $r_4 = (1.5, 0, \alpha_3)$ and b) Longitudinal, i.e., $r_1 = (0, 0, 0)$, $r_2 = (0, 0.5, \frac{\pi}{3})$, $r_3 = (0, 1, \frac{\pi}{3}), r_4 = (0, 1.5, \alpha_3)$ of the tubular membrane.. $\alpha_3$ is varied from 0 to $\frac{9\pi}{10}$.}
  \label{4dpdis}
\end{figure}

\newpage 
where $\left(\nabla^{cyl} \times  \bm{v}^{dipole}\right) $ denotes the curl of $ \bm{v}^{dipole}$ in cylindrical co-ordinates and the dipole velocity field $ \bm{v}^{dipole}$ is given by Eq.~(\ref{vdp}).
%\textbf{Adding the repulsive force}:
%To add the repulsive force, we need to first define a piecewise function $\bm{\psi}_{rep}'$ as follows (the notation has a prime because it is derivative of a potential function) :
%\beqa
%\bm{\psi}_{rep}^{\prime} [x]= \begin {cases}  U_0 ~( 1 - \frac{2 r_s}{x}) & x \leq 2 r_s\\
%0 &   x > 2 r_s  \end {cases}
%\eeqa
%The parameter $r_s$ will be the size of the dipole and the parameter $U_0$ will control the strength of the repulsive force. We will choose $U_0=0.05$ and $r_s=0.075$ (may need to tune them later). The dynamical equations with both hydrodynamic and repulsive forces now become
%
%Dynamical Equations :
%\beqa
%&R ~ \dot{\theta}_i=  \sum_{j \neq i}^N  \bm{v}^{dipole}_{\theta}[\theta_i,z_i,\alpha_i,\theta_j,z_j,\alpha_j] - \bm{\psi}_{rep}^{\prime} [\gamma_{ij}] R~ (\theta_i -\theta_j)\nn\\
%&\dot{z}_i=  \sum_{j \neq i}^N   \bm{v}^{dipole}_{z}[\theta_i,z_i,\alpha_i,\theta_j,z_j,\alpha_j]  \bm{\psi}_{rep}^{\prime} [\gamma_{ij}]~ (z_i -z_j)\nn\\
%&\dot{\alpha}_i=  \sum_{j \neq i}^N \underbrace{\frac{1}{2} \left(\nabla^{cyl}_i \times \bm{v}^{dipole}[\theta_i,z_i,\alpha_i,\theta_j,z_j,\alpha_j]\right).\hat{r}_i}_{\text{Vorticity induced rotation}} 
%\label{dynmeqcyl}
%\eeqa
%where $\gamma_{ij} =\sqrt{(z_i -z_j)^2 + R^2 (\theta_i- \theta_j)^2}$.\\\\
All the simulations are performed in python using standard numerical routines setting $\eta_{2D} =1, \eta=0.01, R=1$ for which $\frac{\lambda}{R} =100$. In the remainder of this section, we will combine the co-ordinates  of the i-th dipole $(\theta_i, z_i)$ on the cylinder surface and its orientation  $\alpha_i$ with respect to $\hat{\theta}$, into a single row vector $ \vec{r}=(\theta_i, z_i, \alpha_i)$. Fig.~(\ref{3d}) illustrates six interesting situations for the dynamics of a pair of dipoles in the tubular fluid membrane. Zoomed versions of the same figures showing the temporal variations in dipole co-ordinates are provided in Fig.~(\ref{zoomedfig}). The temporal evolution of orientation of dipole pair and the corresponding trajectories in $\theta-z$ plane are presented in Fig.~(\ref{azoomedfig}) of Appendix \ref{amap}.
We now explain the six scenarios considered in Fig.~(\ref{3d}) (with the zoomed trajectories displayed in Fig.~(\ref{zoomedfig})): 
\begin{itemize}
\item Case A: Here we observe that a mutually perpendicular dipole pair initially situated along the angular $\hat{\theta}$ direction of the tubular membrane move together along the circumference without any change in orientation and without any deviation along the $\hat{z}$ direction. From Eq.~(\ref{dynmeq}) and  Eq.~(\ref{zlim}) in Appendix \ref{aderiv}, we see that for both dipoles $\dot{\alpha}$ and $\dot{z}$ vanish in this case, in agreement with our numerical observations. Furthermore, since $\dot{\theta}_1 = \dot{\theta_2}$, the dipoles move with a fixed separation.  Note that the fixed point of the system in this situation is given by $z_1 = z_2,~ \theta_1 - \theta_2 = (2n+1)\pi, n \in \mathbb{Z}$.
\item Case B: This situation is similar to Case A, now the mutually perpendicular dipole pair is initially situated along the axial $\hat{z}$ direction of the cylinder. Here we observe that the dipole pair move together along the axis of the cylinder, with no change in dipole separation or orientation. This can also be seen mathematically from  Eq.~(\ref{dynmeq}) and  Eq.~(\ref{thetalim}) in Appendix \ref{aderiv}. Moreover, we observe that the speed of the dipoles in this situation is the same as in Case A.
\item Case C: This situation is same as in Case A, except now the dipoles are not mutually perpendicular. In this situation, the dipole pair still moves along the $\hat{\theta}$ direction but with non-linear oscillations in their orientations, as observed in Case C of Fig.~(\ref{azoomedfig}) in Appendix \ref{amap}. However, as evident from Eq.~(\ref{dynmeq}) and  Eq.~(\ref{zlim}) in Appendix \ref{aderiv}, the dipole trajectories remain strictly confined in the $\hat{\theta}$ direction, with no deviation along the $\hat{z}$ direction. The separation between the dipoles oscillates  with time as shown in Fig.~(\ref{to_do}).

\item Case D: This situation is similar to Case C, where the dipoles are not mutually perpendicular. However now the dipoles are initially situated along the $\hat{z}$ direction of the cylinder. Since the cylindrical geometry breaks the in-plane rotational symmetry of the membrane, we observe an interesting effect. The dipole pair moves together along the $\hat{z}$ direction but with a finite gradual ``drift" along the $\hat{\theta}$ direction, clearly visible from the dipole trajectories in Case D of Fig.~(\ref{3d}). This is remarkably different from membranes which preserve the in-plane rotational symmetry, such as flat and spherical membranes. A careful examination of the angular ($\theta$) variations of the dipoles in the zoomed plots in Fig.~(\ref{zoomedfig}) reveals that the dipole motion along the $\hat{\theta}$ direction is a combination of nonlinear oscillations and a finite drift that increases \textit{linearly} with time. We choose to measure the drift $\equiv \theta_D (t)$ (ie. drift at a particular time t) as the slope of the mean motion ie. $\frac{\theta_1(t) +\theta_2(t)}{2}$ along the $\hat{\theta}$ direction (after removing the oscillatory component) multiplied by the elapsed time t. Tracing back to the dynamical equations (Eqs.~(\ref{dynmeq})) it is easy to see that the drift we observe is a consequence of the characteristic local \textit{rigid} rotation along the compact angular direction of the tubular fluid membrane. We  plot the drift   against various parameters. Fig.~(\ref{drift_fig}) shows that $\theta_D (t)$  increases linearly with time. The variation of the drift as a function of the initial relative orientation  $\alpha_I$ is well approximated by a sinusoidal curve with a nonlinear prefactor.
\item Case E: Here we consider a mutually perpendicular dipole pair situated along some arbitrary direction which makes a non-zero angle with respect to $\hat{\theta}$. In all such situations, the dipoles move together along a helix as illustrated in Fig.~(\ref{3d}). Note that the dipoles move in a helix centered around $\hat{z}$, with finite non-linear oscillations in the orientation, see Case E of Fig.~(\ref{azoomedfig}) in Appendix \ref{amap}. 
\item Case F: Here we simulate a two-dipole system, with the same configuration as Case D but the dipole strengths are now opposite such that we have a pusher and a puller type dipole respectively. Notably we also get a non-zero drift in this situation arising from the local \textit{rigid} rotation, see Case F of Fig.~(\ref{3d}) and Fig.~(\ref{zoomedfig}) for illustrations.
\end {itemize} 
\begin{figure}[h]
  \centering
  \includegraphics[width=.35\linewidth]{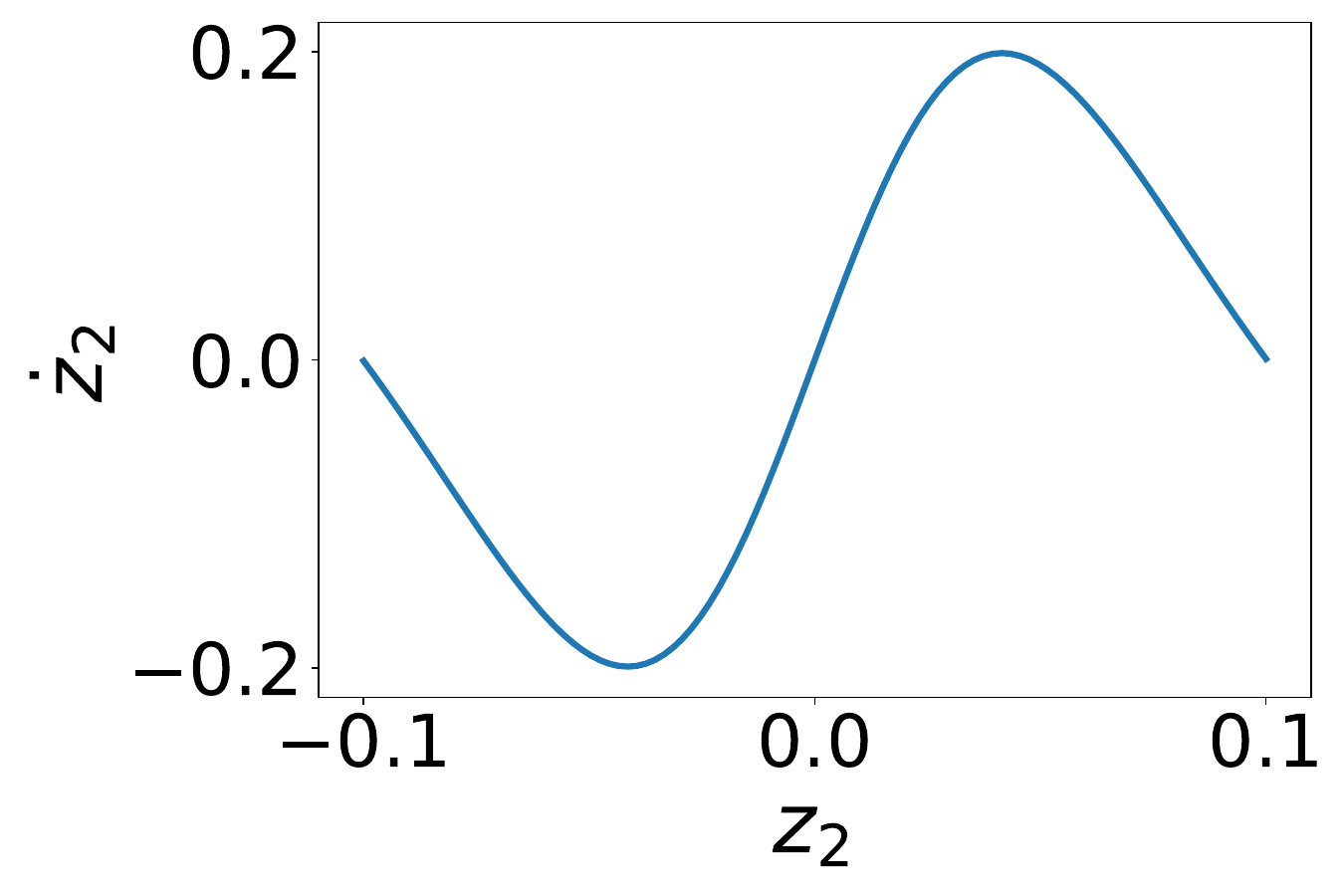}
  
  \caption{Plot of $\dot{z}_2$ against a small perturbation $z_2$ in Case A.}
  \label{case_a_stability}
\end{figure}
It is interesting to note that the only closed orbit possible for two dipole systems is the one obtained in Case A i.e. the orbit covers the entire circumference of the tubular membrane (full orbit). This can be understood as follows: let there be a closed orbit in the $\theta-z$ plane other than the full orbit (Case A). For such an orbit to exist, at some point in time the $z$ coordinates of the dipoles must be equal. However, we note from Eq.~(\ref{dynmeq}) and Eq.~(\ref{zlim}) of Appendix \ref{aderiv} that for $z \rightarrow z_0,~ v_z = 0$ irrespective of $\theta$ and $\alpha$. This implies that the only closed orbit possible is the ring that covers the entire circumference of the tubular membrane. We can analyze the stability of the orbit in case A numerically. In Fig.~(\ref{case_a_stability}), we plot $\dot{z}_2~(t = 0)$ for the initial conditions $r_1 = (0, 0, 0)$ and $r_2 = (0.1, z_2(0),\frac{\pi}{2})$, where $z_2(0)$ is a small perturbation. We observe that the perturbed dipole is pushed out further in the direction of the perturbation. Thus we conclude that the orbit is unstable. We also study the temporal evolution of the inter-dipole separation squared $S(t) \equiv R^2 ~\text{Min} \left[\theta_1 -\theta_2, 2\pi -(\theta_1 -\theta_2) \right]^2 +(z_1-z_2)^2 $ which is a distance function appropriate for the cylinder geometry. For example, in Case C,  we observe that the inter-dipole separation oscillates with a time period that scales as the square of the initial separation, for small separations, see Fig.~(\ref{to_do}). In addition to the six interesting situations discussed in this section (Case A-F), a more elaborate map of possible two-dipole trajectories with respect to initial location and orientation are presented in Appendix \ref{amap}. \\\\
\textbf{Aspects of Multi-dipole configurations}: The number of tunable initial parameters rapidly proliferates for more than two dipoles. Here we choose to investigate the impact of cylindrical geometry of the membrane using two quantities : the drift $\theta_D$ and the sum of squared inter-dipole distances. Based on the results of the two-dipole system, one expects that the loss of in-plane rotational symmetry of the cylindrical membrane should lead to a drift in multi-dipole systems as well. We observe that this is indeed true. For example, we observe a non-zero drift for three and four dipole systems. To illustrate this, we choose an initial configuration of three or four dipoles along the longitudinal direction of the tube with random initial orientations. The resulting dynamics shows a finite drift along the angular direction of the tube, see Fig.~(\ref{theta_d multi dipole}), for both three-dipole and four-dipole systems (the drift $\theta_D$ for a multi-dipole system is simply measured by the average of the angular ($\theta$)  co-ordinate of the multi-dipole system, ie. $\sum_i \theta_i /N$ where N is the number of dipoles).\\\\
A second way to investigate the consequences of breaking of in-plane rotational symmetry of the membrane is to study how the inter-dipole distances vary with time, say for a configuration of dipoles placed initially along the transverse direction ($\hat{\theta}$) and another configuration placed along the longitudinal ($\hat{z}$) direction. We define the sum of squared inter-dipole distances for multi-dipole configurations as  
\beqa
S(t)\equiv \sum_{i\neq j}  R^2 ~\text{Min} \left[\theta_i -\theta_j, 2\pi -(\theta_i -\theta_j) \right]^2 +(z_i-z_j)^2 
\label{Sdef}
\eeqa
which is appropriate for the cylindrical membrane geometry.
In Fig.~(\ref{3dpdis}) and Fig.~(\ref{4dpdis}) we study the evolution of $S(t)$ for the three dipole and four dipole system respectively. Focusing on the three dipole system Fig.~(\ref{3dpdis}), we first consider a configuration of three dipoles, all situated along the transverse direction of the tube ie.  $r_1 = (0, 0, 0)$, $r_2 = (0.5, 0, \frac{\pi}{3})$, $r_3 = (1, 0, \alpha_3)$ and study the evolution of $S(t)$ upon varying the orientation $ \alpha_3$ of the rightmost dipole. This is presented in the left panel of Fig.~(\ref{3dpdis}). We contrast this with a similar configuration of three dipoles aligned along the longitudinal direction ie. $r_1 = (0, 0, 0)$, $r_2 = (0, 0.5, \frac{\pi}{3})$, $r_3 = (0, 1, \alpha_3)$. The evolution of $S(t)$ for this situation is shown on the right panel of Fig.~(\ref{3dpdis}). Similar investigation is carried out for four dipoles in Fig.~(\ref{4dpdis}). We observe remarkable differences in the evolution of  $S(t)$. The growth in separation is mostly non-monotonic for configurations that are initially along transverse direction, compared to configurations along longitudinal direction of the tube, demonstrating the impact of breaking of in-plane rotational symmetry of the membrane.

\section{Summary and future directions}
\label{cncl}
We summarize the results obtained in this paper. We construct viscous fluid flow sourced by a force dipole embedded in a tubular fluid membrane coupled to external embedding fluids. We find analytic expressions for the flow in the limit of infinitely long and thin tubular membranes. The flow field exhibits a quadrupolar defect structure surrounding the dipole location and four saddles, such that the net topological index is zero, consistent with the Euler characteristic of the cylinder, as dictated by the Poincar\'e index theorem. The streamlines of the flow field are thus remarkably different from the corresponding flows in spherical \cite{sarthak22} or flat membranes \cite{mnk}. Using the analytic solution, we formulate the dynamics of a pair of pusher-type dipoles. We find that a mutually perpendicular dipole pair generically moves together along helical geodesics along the cylinder surface. An interesting aspect of the cylindrical geometry is that it breaks the in-plane rotational symmetry of the membrane. This leads to a major difference in dipole flows along the axial and transverse directions of the tube, which in turn leads to anisotropic hydrodynamic interaction between the dipoles. The dynamics are thus distinct from flat and spherical membranes. In particular, the flow along the angular direction has a \textit{local} rigid rotation term. Due to this feature of the flow, we observe that the interacting dipole pair initially situated along the axis of the cylinder exhibits an overall drift along the compact angular direction of the cylinder. Focusing on the two-dipole system, we also show that the only possible closed orbit for a two-dipole system is the one encircling the full cylinder around the $\hat{z}$ axis (Case A). In particular, librations are not allowed.  Our results are relevant for non-equilibrium dynamics of motor proteins in tubular membranes arising in nature, as well as in-vitro experiments, particularly the interesting and beautiful work  \cite{bss11}. Moreover, the dynamics of motor proteins in fluid membranes is now an active area of experimental studies \cite{aggexp1,aggexp2}.\\\\
The model we have considered here offers a course-grained fluid-mechanics perspective on a wide class of systems, featuring a viscous membrane tube separating external embedding fluids of different viscosities and hosting mobile inclusions. For example, the diffusion experiments performed in tubular membranes (\cite{bss11}) showed good agreement with Saffman-Delbr\"uck theory.  Motor proteins typically exert forces of  $\mathcal{O}$(pico-Newton), and the Saffman length is of $\mathcal{O}$($\mu$m). This induces membrane flows of  $\mathcal{O}$($\mu$m/s). The topological aspects of the streamlines reported in this work may also be probed via accurate particle image velocimetry (PIV) along the lines of \cite{wg2013}. We hope our results will motivate more experimental investigations on the dynamics of motor proteins in membrane tubes similar to \cite{bss11} or along the lines of \cite{ckst2022}, where hydrodynamically induced helical particle drift was recently reported in patterned surfaces.\\\\
We end with brief comments on subsequent works that we wish to report in the future.  In this paper, our focus has been on force dipole interactions in tubular fluid membranes. It will be interesting to extend these results to a large collection of interacting dipoles in the tubular membrane (since membranes in nature typically host a large number of inclusions). The dynamics of a mixture of pusher and puller type motors interacting in a tubular geometry will also be very interesting. Another direction worth exploring is to generalize these results to situations where the
external fluid is confined via a substrate. It will also be important and interesting to include thermal fluctuations in this setup following the work of Sokolov and Diamant \cite{sokolov_haim2018}. Detailed investigations of rotating proteins in tubular fluid membranes will be covered in an upcoming manuscript  \cite{rs23}. 
\section{Acknowledgements}
R.S. acknowledges support from DST INSPIRE, India (Grant No.IFA19-PH231) and the OPERA Research Grant from Birla Institute of Technology and Science, Pilani (Hyderabad Campus). We are very thankful to Anantha Aiyyer, Brandon Clarke, Mark Henle,  Prasad Perlekar, Mustansir Barma, Sarthak Bagaria, Ishan Mata and Joe Ninan for discussions and suggestions. R.S thanks  TIFR Hyderabad for providing the opportunity to present parts of this work. S.J. would like to thank Sehej Jain for relevant discussions on the topic.
\appendix
\section{Construction of Force Dipole Solution}
\label{aderiv}
In this Appendix, we derive Eq.~(\ref{vstkhc}) of the main text, following Ref.~\cite{henlev2008} and Ref.~\cite{sarthak22}.
 We need to solve the system of equations described by Eq.~(\ref{systemeq}) for the velocity field $v_\alpha$  in response to an external stress  $\sigma^{ext}_{\alpha}$ generated by a point force dipole. Since we are interested in a tubular fluid membrane, the relevant 2D membrane metric in cylindrical co-ordinates $x^\mu=(\theta,z)$ is given by $ds^2 = R^2 d\theta^2 +dz^2$ and we set the local curvature $K(x) =0$ in  Eq.~(\ref{systemeq}). The solution requires lengthy but straightforward algebraic manipulations, carried out in \textit{Mathematica} \cite{wolfram} .\\[5pt]
\textbf{ Solution ansatz for the membrane fluid flow:} We will consider the following ansatz for the velocity field of the membrane fluid 
\beqa
v_{\alpha}(\theta, z)=i ~\epsilon_{\alpha \beta} \sum\limits_{n=-\infty}^{\infty}\int\limits_{-\infty}^{\infty} dq ~A(\Lambda) ~ \Lambda^{\beta} e^{i \Lambda_\mu x^{\mu}},
\eeqa
%\Phi(\vec{x} ; \Lambda)=\exp \left[e^{i \Lambda^{x^{\mu}}}\right], \quad \lambda(\Lambda)=-\eta_{\mathrm{m}} \Lambda_{\alpha} \Lambda^{\alpha}
%$$
%where the covariant vector $\Lambda_{\alpha}$ is defined as
%$$
%\Lambda_{\theta}=n, \quad \Lambda_{z}=q, \quad \Rightarrow \Lambda_{\alpha} \Lambda^{\alpha}=\left[q^{2}+\frac{n^{2}}{R^{2}}\right]
%$$
where we have adopted the shorthand  notation  $\Lambda$  for the Fourier modes on the cylinder with co-ordinates $x^\mu=(\theta,z)$
$$
\Lambda_{\theta}=n, \quad \Lambda_{z}=q, \quad \Rightarrow g^{\alpha \beta}\Lambda_{\alpha} \Lambda_{\beta}=\left[q^{2}+\frac{n^{2}}{R^{2}}\right]
$$
\textbf{ Solution for 3D flows in external embedding fluids:} Following Happel and Brenner \cite{hb}, the solution to the Stokes equations for the external 3D embedding fluids have the generic form
\beqa
&\vec{v}^{\pm}(r, \theta, z)= \vec{\nabla} f^{\pm}(r, \theta, z)+\vec{\nabla} \times[g^{\pm}(r, \theta, z) \hat{z}] 
+r \partial_{r}\left[\vec{\nabla} h^{\pm}(r, \theta, z)\right]+\partial_{z} h^{\pm}(r, \theta, z) \hat{z}, \nn\\
&p^{\pm}(r, \theta, z)=-2 \eta_{\pm} \partial_{z}^{2} h^{\pm}(r, \theta, z),\nn
\eeqa

where $f^{\pm}, g^{\pm},h^{\pm}$ are harmonic functions for the 3D Laplacian in cylindrical co-ordinates. They have the generic decomposition

$$
\Psi^{\pm}(r, \theta, z)=\sum_{n=-\infty}^{\infty}\int_{-\infty}^{\infty} dq~  e^{i \Lambda_\mu x^{\mu}} \Psi^{\pm}(\Lambda)~ \xi_{n}^{\pm}(q r)
$$
where
$$
\xi_{n}^{+}(q r) \equiv K_{n}(|q| r), \quad \xi_{n}^{-}(q r) \equiv I_{n}(|q| r)
$$
where $K_n$ and $I_n$ are modified Bessel functions of order $n$ of first and second kind respectively. The covariant components of the 3D external solvent flows can be cast as follows:\\[20pt]
$w_{r}^{\pm}(\Lambda, r)=|q| \widetilde{\xi}_{n}^{\pm}(q r) F^{\pm}(\Lambda)+\frac{i n}{r} \xi_{n}^{\pm}(q r) G^{\pm}(\Lambda)$ $+H^{\pm}(\Lambda)\left[-|q| \widetilde{\xi}_{n}^{\pm}(q r)+r\left(q^{2}+\frac{n^{2}}{r^{2}}\right) \xi_{n}^{\pm}(q r)\right]$\\
$w_{\theta}^{\pm}(\Lambda, r)=i m \xi_{n}^{\pm}(q r) F^{\pm}(\Lambda)-|q| r \widetilde{\xi}_{n}^{\pm}(q r) G^{\pm}(\Lambda)$ $+i n H^{\pm}(\Lambda)\left[|q| r \widetilde{\xi}_{n}^{\pm}(q r)-\xi_{n}^{\pm}(q r)\right]$\\
$w_{z}^{\pm}(\Lambda, r)=i q \xi_{n}^{\pm}(q r) F^{\pm}(\Lambda)+i q H^{\pm}(\Lambda)\left[|q| r \tilde{\xi}_{n}^{\pm}(q r)+\xi_{n}^{\pm}(q r)\right] .$\\[5pt]

where $\left.\widetilde{\xi}^{+}(q r) \equiv \frac{d K_{n}(u)}{d u}\right|_{u=|q| r},\left.\quad \widetilde{\xi}(q r) \equiv \frac{d I_{n}(u)}{d u}\right|_{u=|q| r}$\\[5pt]
\textbf{No-slip boundary conditions:} The no-slip boundary conditions amount to the following 6 equations which are to be solved for the 6 coefficients $F^{\pm}, G^{\pm},H^{\pm}$ in terms of membrane mode coefficient $A$.
\beqa
&w_{r}^{\pm}(\Lambda, R)=0, \quad \Lambda^{\alpha} w_{\alpha}^{\pm}(\Lambda, R)=0 \nn\\
&i \epsilon^{\alpha \gamma} \Lambda_{\gamma} w_{\alpha}^{\pm}(\Lambda, R)=-\Lambda_{\beta} \Lambda^{\beta} A(\Lambda)\nn
\eeqa
The solution for the 6 coefficients $F^{\pm}, G^{\pm},H^{\pm}$ involves lengthy expressions that we solve using \textit{Mathematica}. We now solve the stress balance condition in the second row of Eq.~(\ref{systemeq}). For the LHS of the stress balance equation, we first find the flow sourced by a Stokeslet situated at $(\theta_0,z_0)$. Let us note that the Force Dipole flow will be related to the Stokeslet flow by a directional derivative.
\begin{equation}
\sigma_{\alpha}^{\mathrm{ext}}=\frac{F_{0, \alpha}}{4 \pi^{2} R} \sum\limits_{n=-\infty}^{\infty}\int\limits_{-\infty}^{\infty} dq \exp \left[i \Lambda_{\mu} x^{\mu}-i\left(n \theta_{0}+q z_{0}\right)\right]
\end{equation}

In order to eliminate the membrane pressure, we take an anti-symmetric derivative of the entire stress balance equation and use the following results in order to compute the traction vector $T_\alpha$:\\[5pt]
\beqa
\left.\epsilon^{\alpha \gamma} D_{\gamma} \sigma_{\alpha r}^{\pm}\right|_{r=R}=\eta_{\pm} \int D \Lambda e^{i \Lambda_{\mu} x^{\mu}} \frac{A(\Lambda)}{R^{3}} C^{\pm}(\Lambda)
\eeqa
where 
\beqa
C^{\pm}(n,k)=\frac{2 n^{2}\left[\rho^{\pm}(n,k)\right]^{3}+\left(n^{2}+k^{2}\right)^{2}\left[\rho^{\pm}(n,k)\right]^{2}+2 \rho^{\pm}(n,k)\left(k^{4}-n^{4}\right)-\left(k^{2}+n^{2}\right)^{3}}{\rho^{\pm}(n,k)~ k^{2}-\left[\rho^{\pm}(n,k)-n\right]\left[\rho^{\pm}(n,k)+n\right]\left[\rho^{\pm}(n,k)+2\right]}
\eeqa
and
\begin{equation}
k \equiv q R . \quad \rho^{\pm}(n,k) \equiv \frac{|k| \tilde{\xi}_{n}^{\pm}(k)}{\xi_{n}^{\pm}(k)}.
\end{equation}
More explicitly,
\beqa
&\rho_+ [n,k] = \frac{|k| \frac{d  K_n[u]}{du} \big{|}_{u= |k|}}{K_n[|k|]}, \hspace{1cm} \rho_- [n,k] = \frac{|k| \frac{d  I_n[u]}{du} \big{|}_{u= |k|}}{I_n[|k|]}.
\eeqa

The solution for $A$ is given by
\beqa
A_{n}(k)=-\frac{i R^{2} e^{-i\left(n \theta_{0}+k \frac{z_{0}}{R}\right)}}{4 \pi^{2} \eta_{2d}~ c_{n}(k)}\left[k\left(\vec{F}_{0} \cdot \hat{\theta}_{0}\right)-n\left(\vec{F}_{0} \cdot \hat{z}\right)\right],
\eeqa
 and
\beqa
c_{n}(k) \equiv\left(k^{2}+n^{2}\right)^{2}+\frac{R}{\lambda_{+}} C_{+}(n,k)-\frac{R}{\lambda_{-}} C_{-}(n,k) .
\eeqa
The final solution for the Stokeslet flow $ \vec{v}^{Stk}(\theta, z)$ for a point force of unit magnitude  $F_0=1$ and making an angle $\alpha_0$ with respect to $\hat{\theta}$ (ie.  $\vec{F}_{0} \cdot \hat{\theta}_{0}=F_0 \cos \alpha_0$ and $ \vec{F}_{0} \cdot \hat{z}=F_0 \sin \alpha_0$)  is given by
\beqa
v^{Stk}_\theta[\theta,z]= G_{\theta \theta} \cos \alpha_0 + G_{\theta z} \sin \alpha_0\nn\\
v^{Stk}_z[\theta,z]=   G_{z \theta} \cos \alpha_0 + G_{z z} \sin \alpha_0,
\label{stk}
\eeqa

\beqa
&G_{\theta \theta} =\frac{1}{4 \pi^2 \eta_{2d}}  \sum\limits_{n=-\infty}^{\infty}~\int\limits_{-\infty}^{\infty} dk~ \frac{k^2}{c_n(k)} e^{i \left( n (\theta-\theta_0) + \frac{k}{R}(z-z_0)\right)}, ~~
G_{\theta z} =-\frac{1}{4 \pi^2 \eta_{2d}}\sum\limits_{n=-\infty}^{\infty}~\int\limits_{-\infty}^{\infty} dk~ \frac{k~ n}{c_n(k)} e^{i \left( n (\theta-\theta_0) + \frac{k}{R}(z-z_0)\right)},\nn\\
&G_{z \theta} = -\frac{1}{4 \pi^2 \eta_{2d}}\sum\limits_{n=-\infty}^{\infty}~\int\limits_{-\infty}^{\infty} dk~ \frac{k ~ n}{c_n(k)} e^{i \left( n (\theta-\theta_0) + \frac{k}{R}(z-z_0)\right)},~~
G_{zz} =\frac{1}{4 \pi^2 \eta_{2d}}\sum\limits_{n=-\infty}^{\infty}~\int\limits_{-\infty}^{\infty} dk~ \frac{n^2}{c_n(k)} e^{i \left( n (\theta-\theta_0) + \frac{k}{R}(z-z_0)\right)},\nn\\
\label{stkgreen}
\eeqa
%The force dipole flow is given by
%\beqa
%v_\theta^{dipole}[\theta,z]= \kappa \left(\frac{\cos \alpha_0}{R} ~ \partial_{\theta_0} + \sin \alpha_0~ \partial_{z_0}\right) v_\theta^{Stk}[\theta-\theta_0,z-z_0] \nn\\
%v_z^{dipole}[\theta,z]= \kappa  \left(\frac{\cos \alpha_0}{R} ~ \partial_{\theta_0} + \sin \alpha_0~ \partial_{z_0}\right) v_z^{Stk}[\theta-\theta_0,z-z_0]
%\label{vdpfull}
%\eeqa
where $\kappa$ is the dipole strength. In the high curvature limit and assuming $\eta_-=\eta_+\equiv \eta$, we have
\begin{equation}
\lim _{R / \lambda_+ \rightarrow 0} c_{n}(k)= \begin{cases}k^{4}+2 k^{2} \frac{R } {\lambda} & n=0 \\[5pt]
 \left(k^{2}+n^{2}\right)^{2} & n \geq 1\end{cases}
\end{equation}
In this limit, we can perform the integrals and sums analytically. For a point force of unit strength $F_0=1$ situated at the origin and making an angle $\alpha_0$ with respect to $\hat{\theta}$ (ie. $\vec{F}_{0} \cdot \hat{\theta}_{0}=\cos \alpha_0$ and $ \vec{F}_{0} \cdot \hat{z}=\sin \alpha_0$), the Stokeslet flow $\bm{v}^{Stk}$ in this limit of high curvature (tubular membrane) is given by
\small{
\begin{align}
&v_\theta^{Stk}[\theta,z]=\frac{1}{8 \pi \eta_{2d} R}  \left(-\frac{z \sin \alpha_0 \sin \theta}{ \cos \theta- \cosh \frac{z}{R}}+\cos \alpha_0 \left(-|z| \frac{e^{-\frac{|z|}{R}} -\cos \theta}{ \cos \theta - \cosh \frac{|z|}{R}}-R~ \log \left[1-2 e^{-\frac{|z|}{R}} \cos \theta+e^{-\frac{2 |z|}{R}}\right]+ \sqrt{2 R \lambda}~ e^{-\frac{\sqrt{2} |z|}{\sqrt{\lambda R}}}\right)\right)\nn\\
& v_z^{Stk}[\theta,z]=\frac{1}{8 \pi  \eta_{2d} R} \left(-\frac{z \cos \alpha_0 \sin \theta}{\cos \theta-\cosh \frac{z}{R}}+\sin \alpha_0 \left(|z| \frac{e^{-\frac{|z|}{R}} -\cos \theta}{ \cos \theta - \cosh \frac{|z|}{R}}-R~ \log \left[1-2 e^{-\frac{|z|}{R}} \cos \theta+e^{-\frac{2 |z|}{R}}\right]\right)\right).
\label{avstkhc}
\end{align}}
The corresponding flow  at the location $(\theta,z)$ sourced by a force dipole of strength $\kappa$, situated situated at $(\theta_0,z_0)$ and making an angle $\alpha_0$ with respect to $\hat{\theta}$ is given by
\beqa
v_\theta^{dipole}[\theta,z]= \kappa \left(\frac{\cos \alpha_0}{R} ~ \partial_{\theta_0} + \sin \alpha_0~ \partial_{z_0}\right) v_\theta^{Stk}[\theta-\theta_0,z-z_0] \nn\\
v_z^{dipole}[\theta,z]= \kappa  \left(\frac{\cos \alpha_0}{R} ~ \partial_{\theta_0} + \sin \alpha_0~ \partial_{z_0}\right) v_z^{Stk}[\theta-\theta_0,z-z_0]
\label{avdp}
\eeqa
where $ v_\theta^{Stk}$ and $ v_z^{Stk}$ are defined in Eq.~(\ref{avstkhc}).\\ Note that in the limit $z\rightarrow z_0$, the force dipole flow Eq.~(\ref{avdp}) simplifies to
\beqa
&\lim _{z\rightarrow z_0}v_\theta^{dipole}= \frac{\kappa \cos 2\alpha_0 ~\cot\left(\frac{\theta- \theta_0}{2}\right)}{8 \pi R~ \eta_{2d}} \nn\\
&\lim _{z\rightarrow z_0} v_z^{dipole}= 0,
\label{zlim}
\eeqa

and in the limit  $\theta\rightarrow \theta_0$, the force dipole flow Eq.~(\ref{avdp}) simplifies to
\beqa
&\lim _{\theta\rightarrow \theta_0}v_\theta^{dipole}= -\frac{ \kappa~(z-z_0) ~ \sin 2\alpha}{4 \sqrt{2} \pi R^{\frac{3}{2}} \lambda^{\frac{1}{2}} \eta_{2d}}+\mathcal{O} \left(\frac{1}{R^{\frac{5}{2}}}\right)\nn\\
&\lim _{\theta\rightarrow \theta_0} v_z^{dipole}= -\frac{\kappa \cos 2\alpha}{4 \pi \eta_{2d} (z-z_0)}+\mathcal{O} \left(\frac{1}{R^2}\right).
\label{thetalim}
\eeqa

\section{Parameter Scan for the Two-dipole System}
\label{amap}
\begin{figure}[h!]

\begin{subfigure}{0.24\textwidth}
\includegraphics[width=\linewidth]{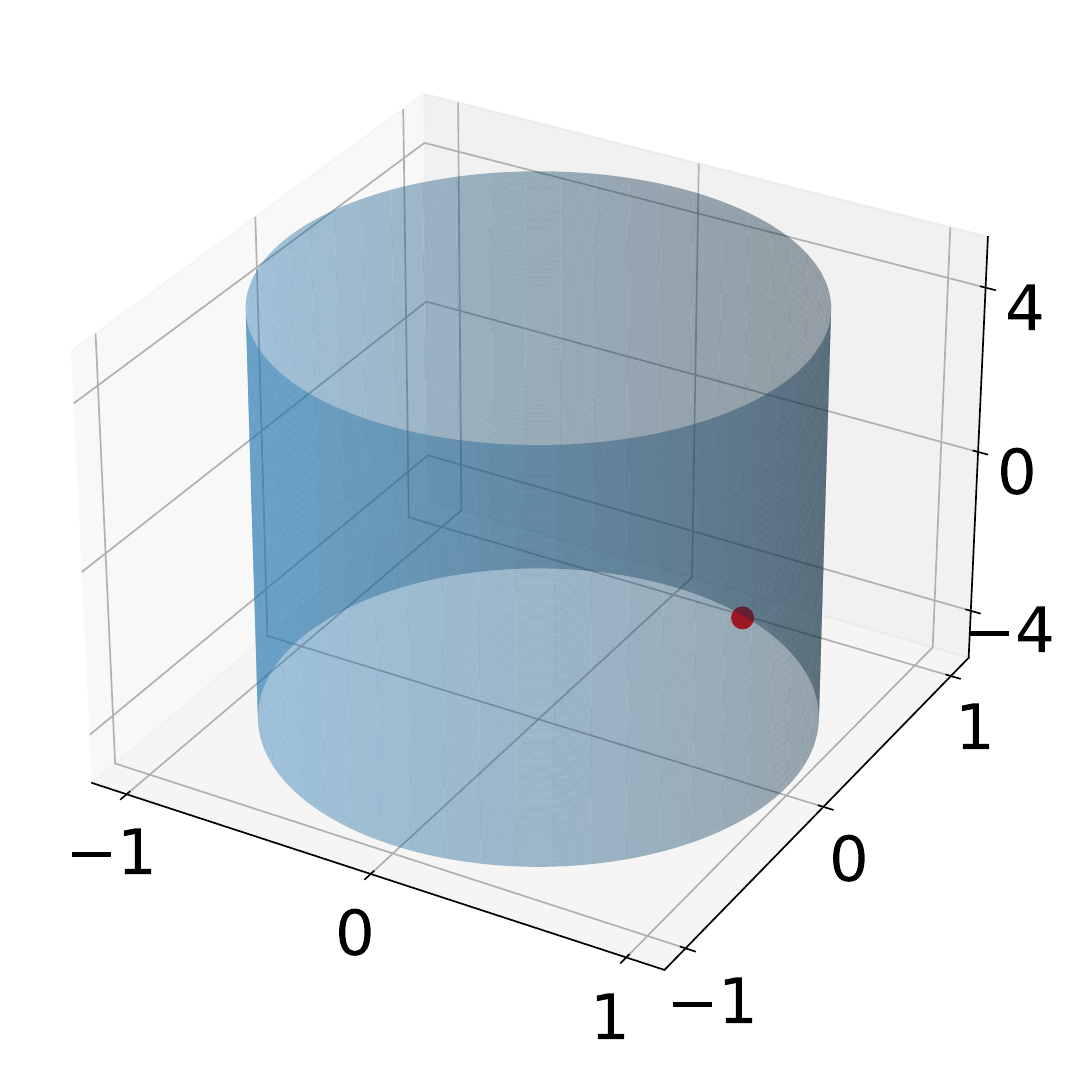}
\caption{$\beta = 0, \alpha_I = 0$}
\end{subfigure} 
\begin{subfigure}{0.24\textwidth}
  \includegraphics[width=\linewidth]{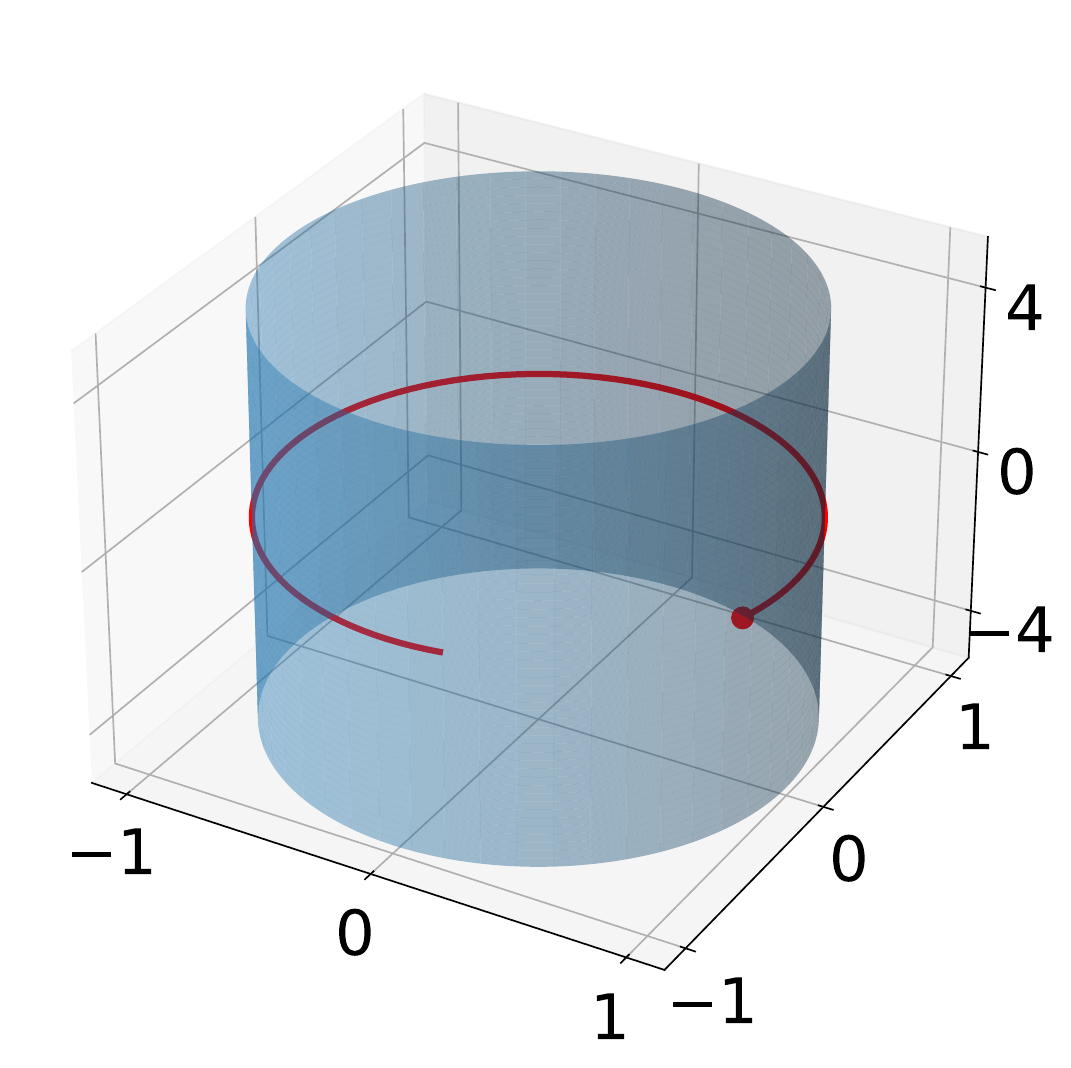}
\caption{$\beta = 0, \alpha_I = \frac{\pi}{6}$ (Case C)}
  \label{matrix_c}
\end{subfigure}
\begin{subfigure}{0.24\textwidth}
\includegraphics[width=\linewidth]{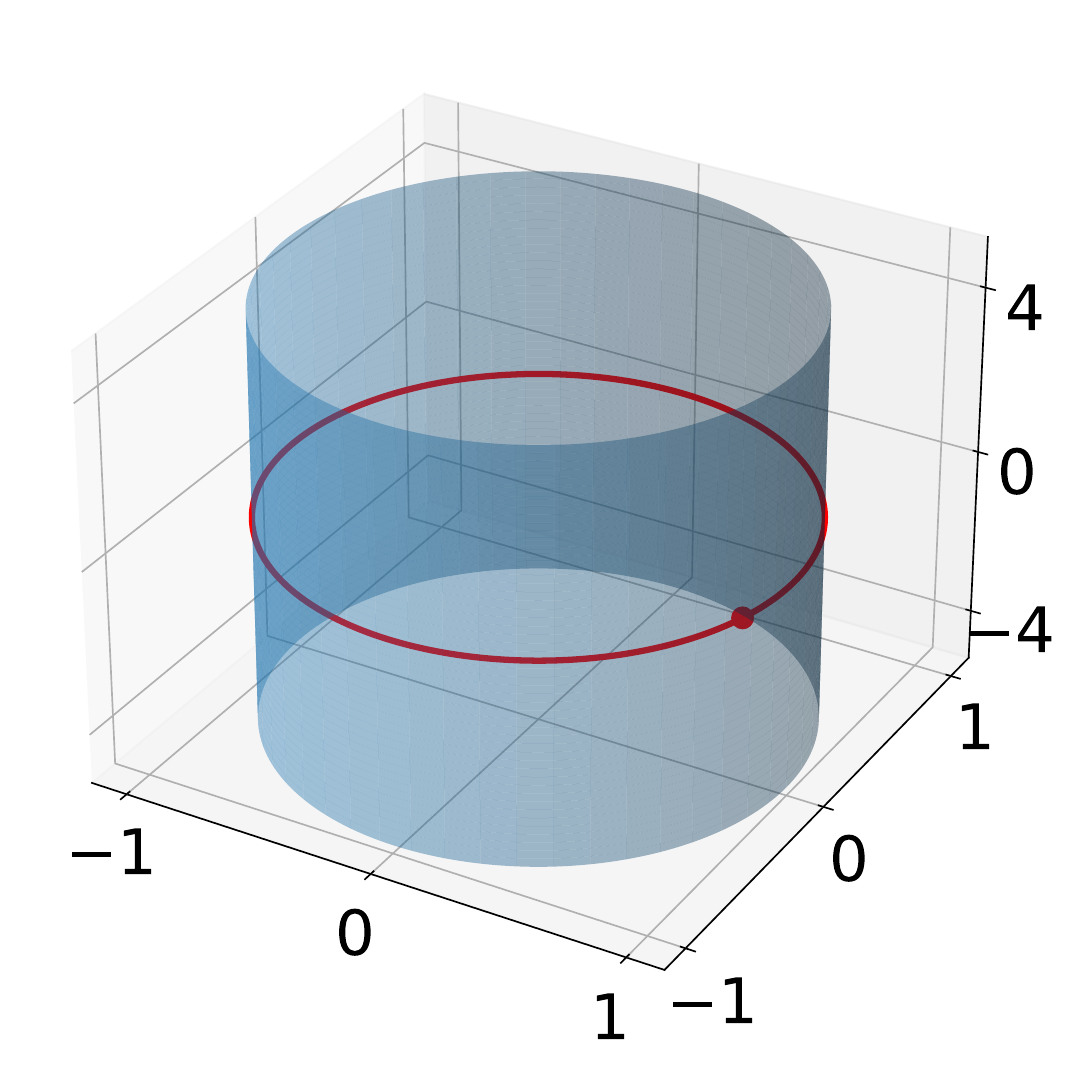}
\caption{$\beta = 0, \alpha_I = \frac{\pi}{3}$
}
\end{subfigure} 
\begin{subfigure}{0.24\textwidth}
  \includegraphics[width=\linewidth]{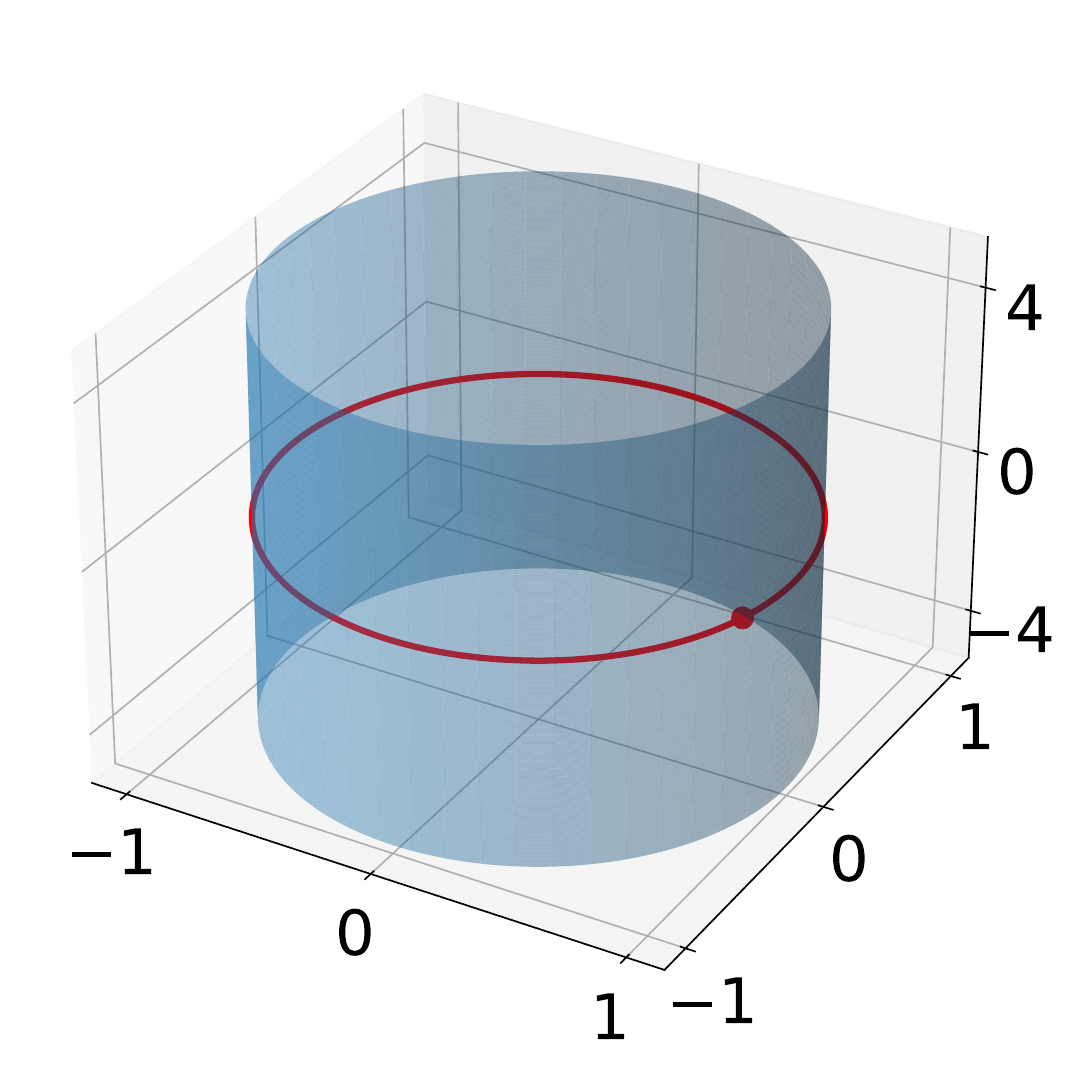}
\caption{$\beta = 0, \alpha_I = \frac{\pi}{2}$ (Case A)
}
  \label{matrix_a}
\end{subfigure}

\begin{subfigure}{0.24\textwidth}
\includegraphics[width=\linewidth]{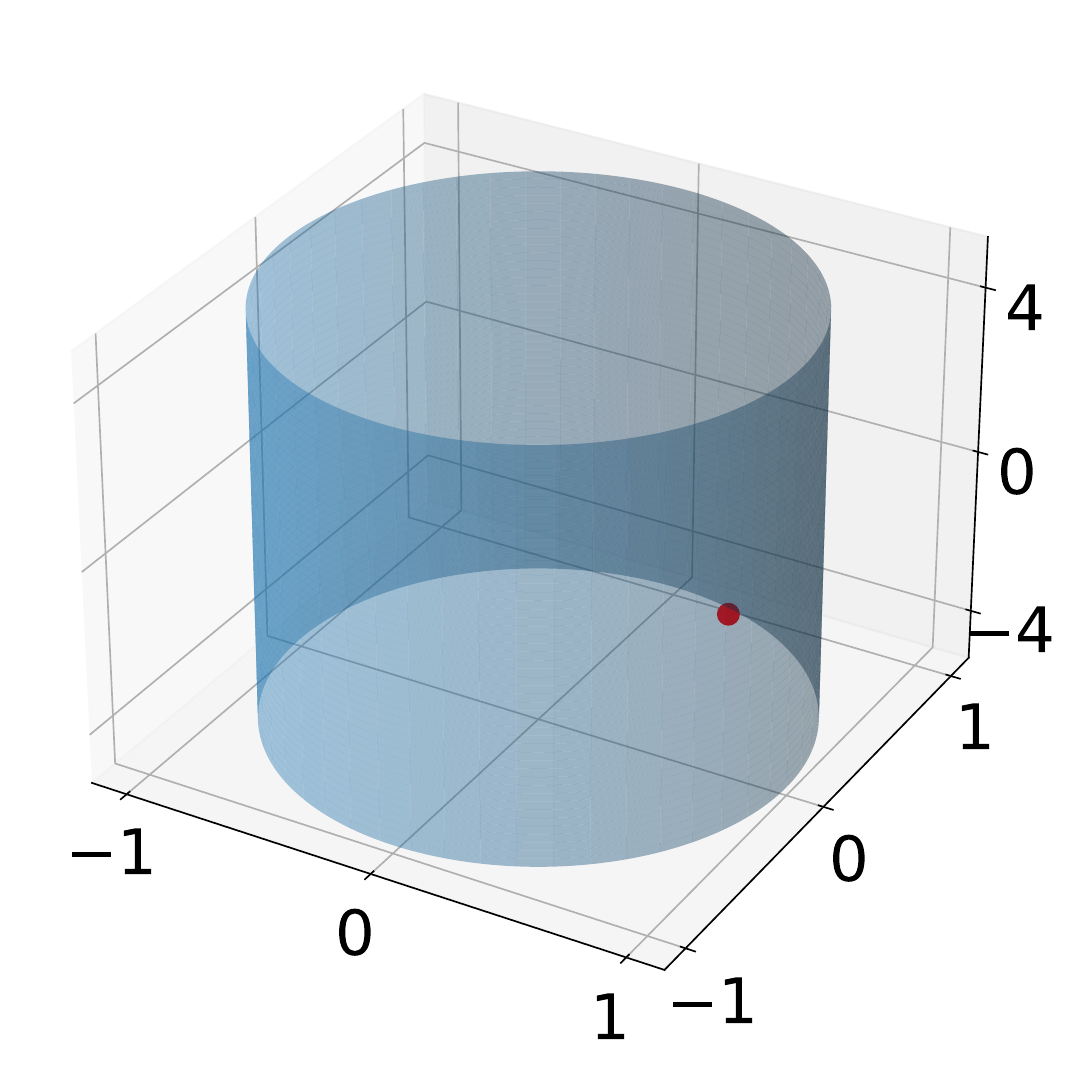}
  \caption{$\beta = \frac{\pi}{6}, \alpha_I = 0$}

\end{subfigure} 
\begin{subfigure}{0.24\textwidth}
\includegraphics[width=\linewidth]{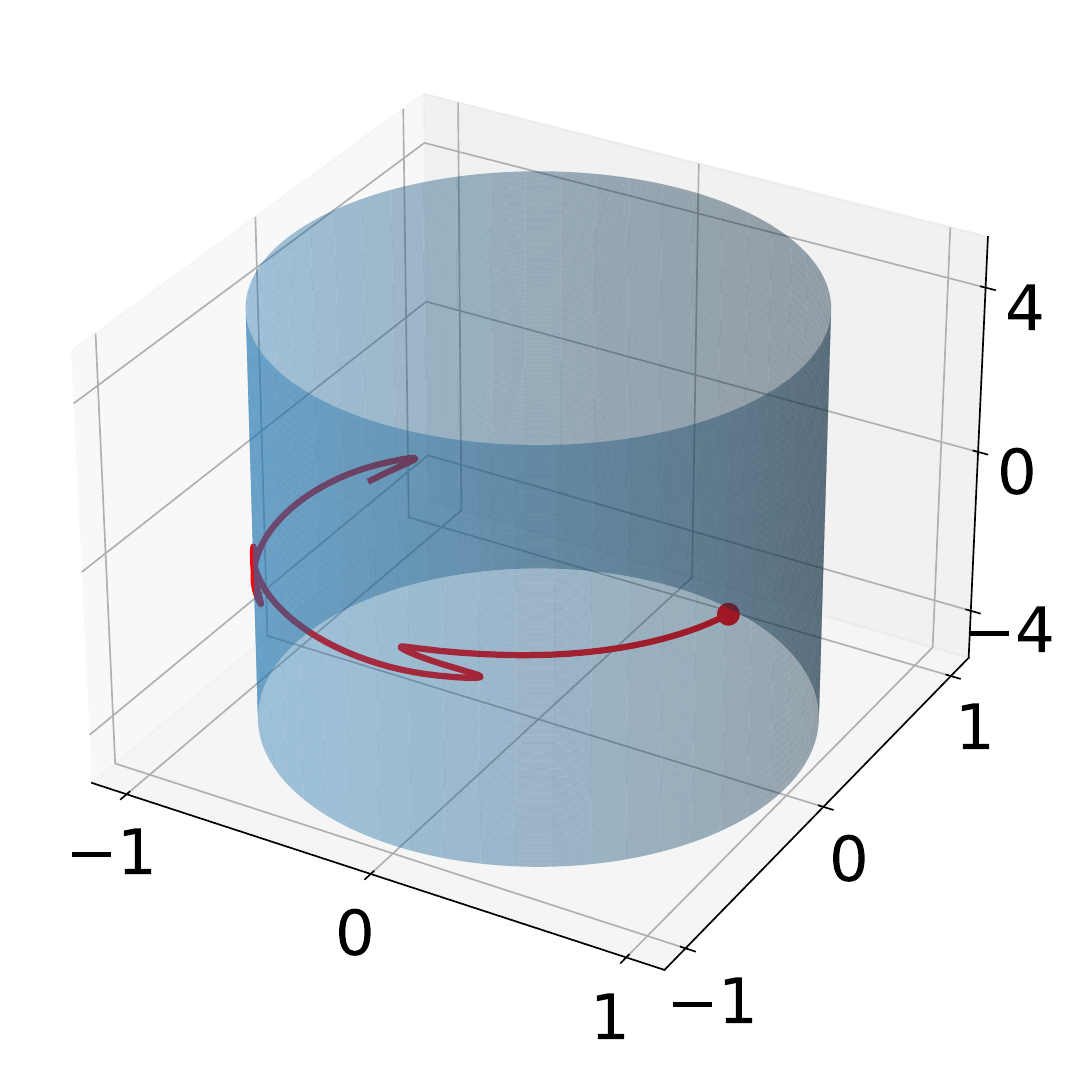}
  \caption{$\beta = \frac{\pi}{6}, \alpha_I = \frac{\pi}{6}$}

\end{subfigure}
\begin{subfigure}{0.24\textwidth}
\includegraphics[width=\linewidth]{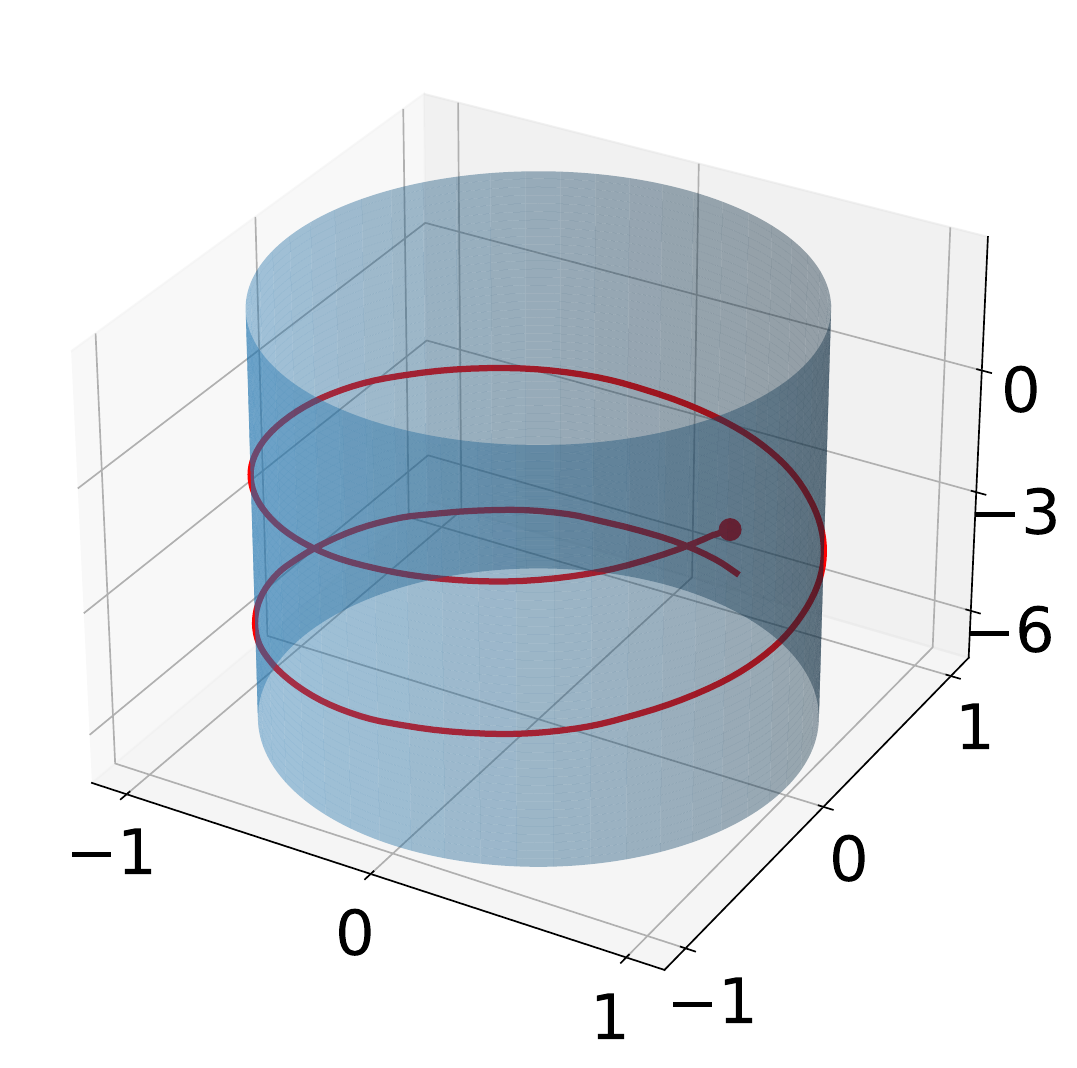}
\caption{$\beta = \frac{\pi}{6}, \alpha_I = \frac{\pi}{3}$}
\end{subfigure} 
\begin{subfigure}{0.24\textwidth}
  \includegraphics[width=\linewidth]{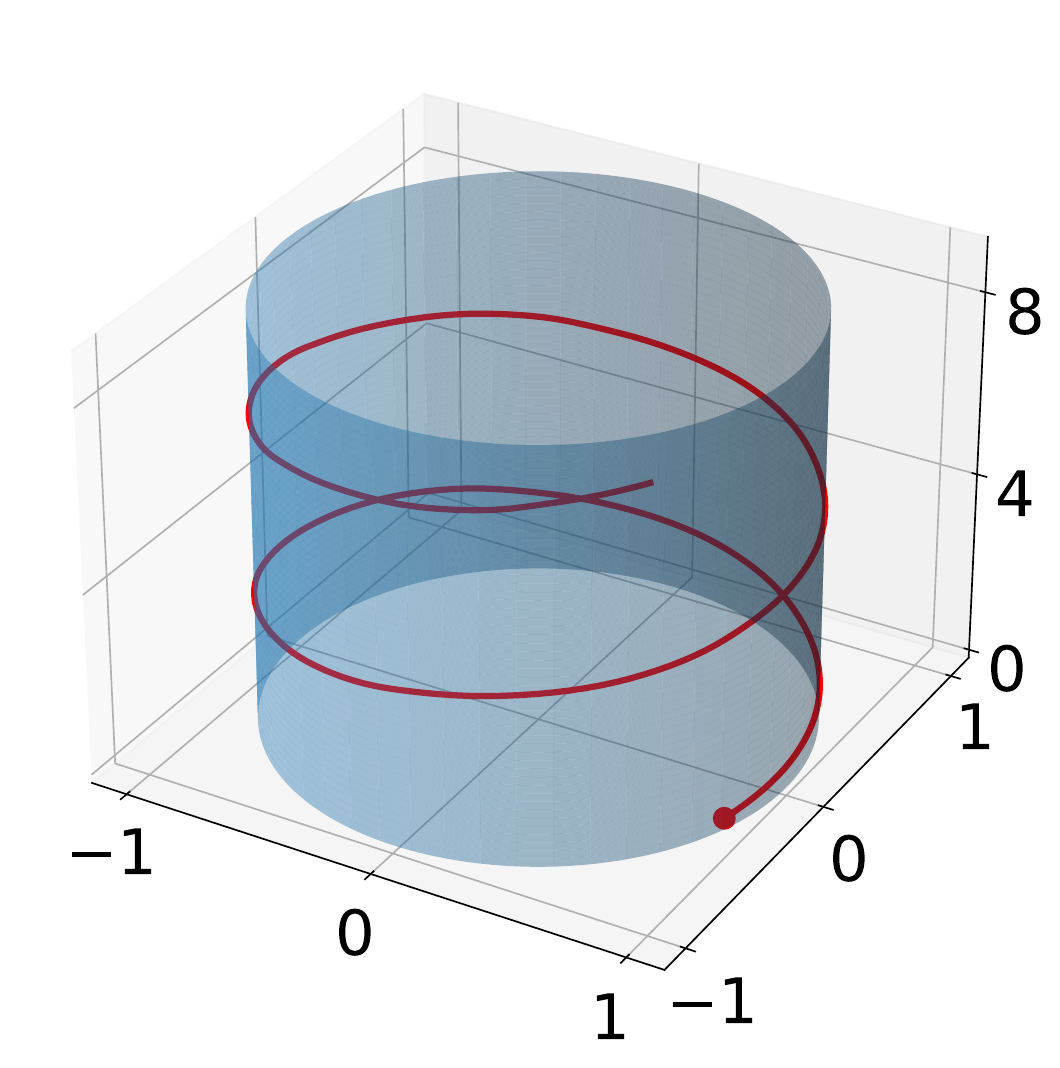}
  
\caption{$\beta = \frac{\pi}{6}, \alpha_I = \frac{\pi}{2}$ (Case E)}
\end{subfigure}
\begin{subfigure}{0.24\textwidth}
\includegraphics[width=\linewidth]{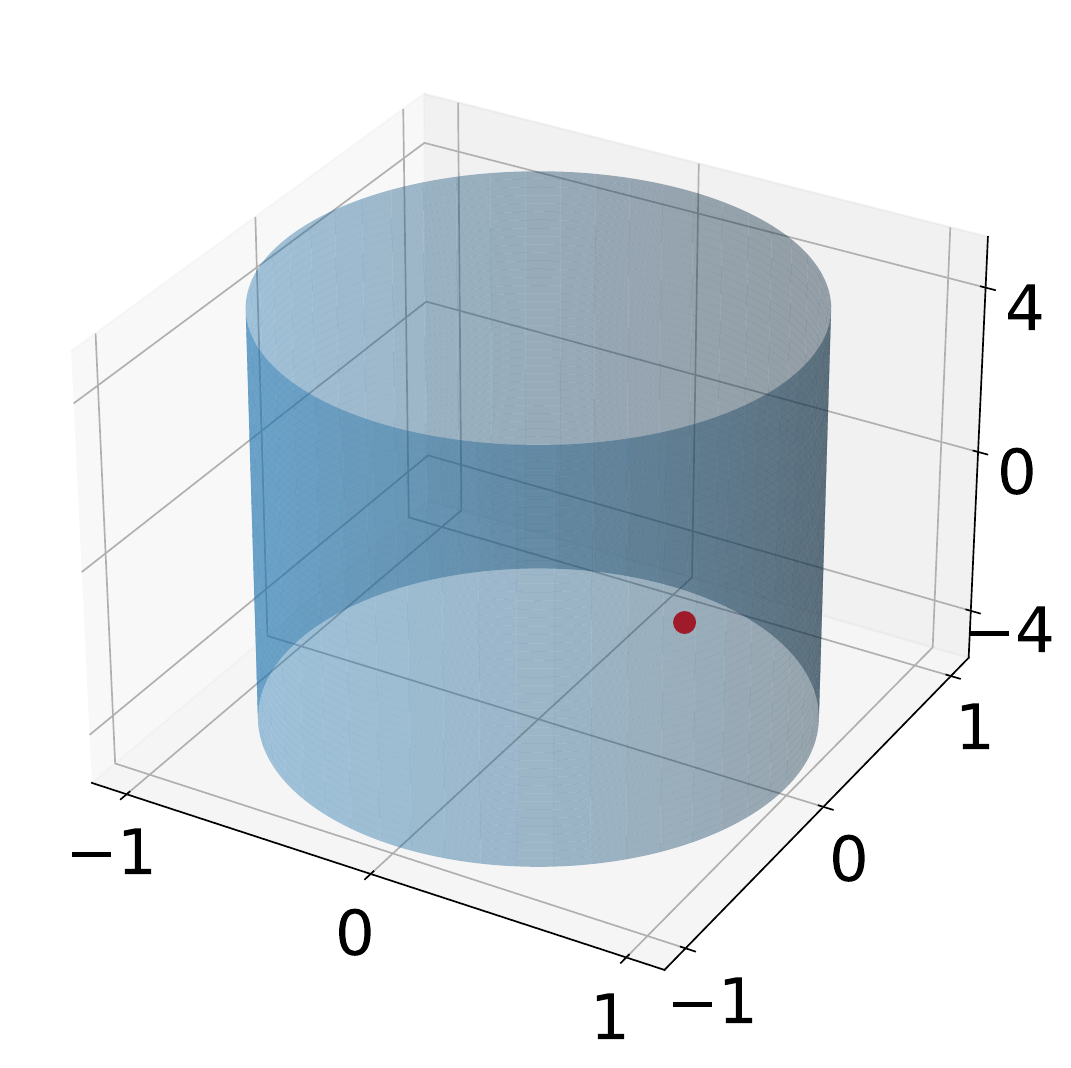}
\caption{$\beta = \frac{\pi}{3}, \alpha_I = 0$}
\end{subfigure} 
\begin{subfigure}{0.24\textwidth}
  \includegraphics[width=\linewidth]{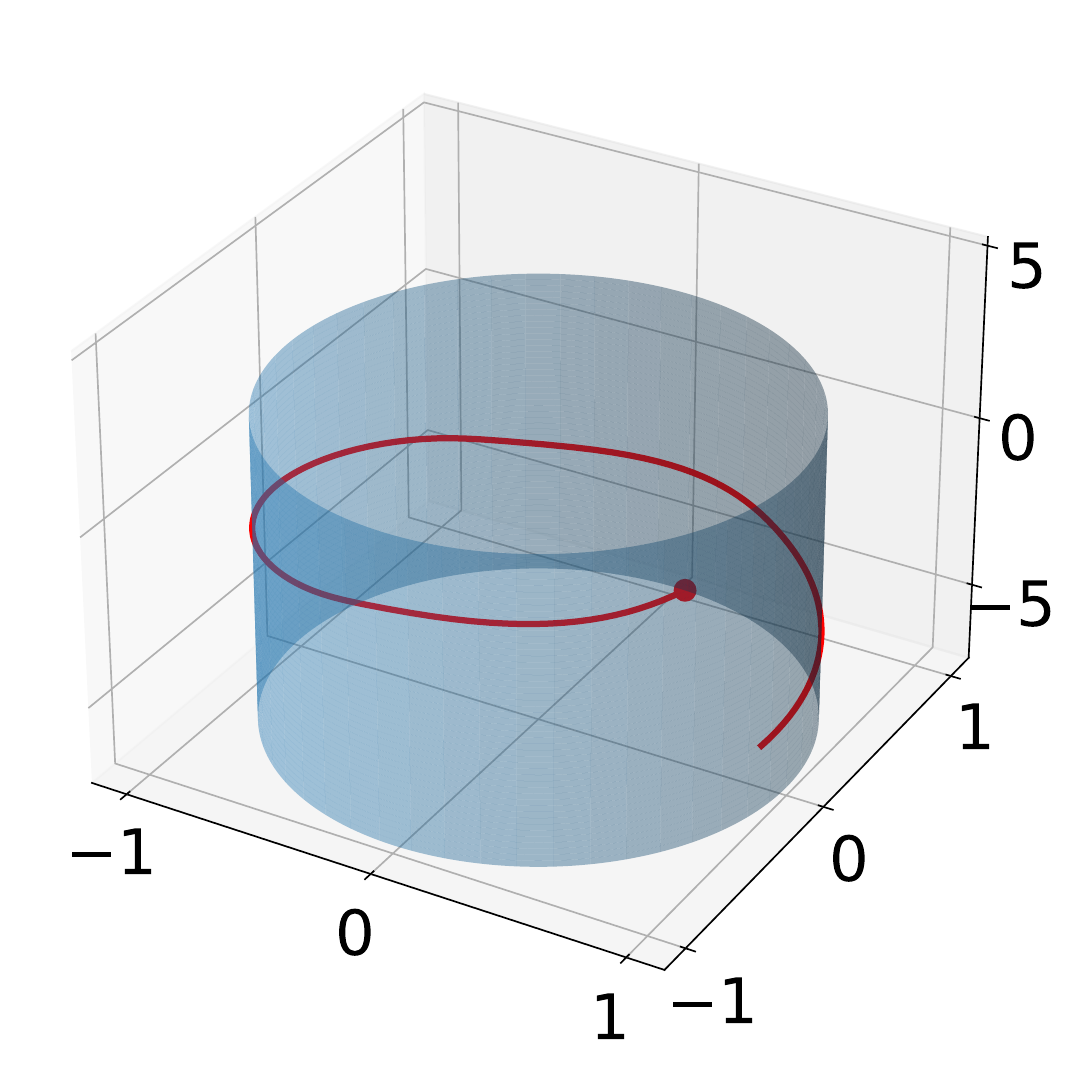}
  
\caption{$\beta = \frac{\pi}{3}, \alpha_I = \frac{\pi}{6}$}
\end{subfigure}
\begin{subfigure}{0.24\textwidth}
\includegraphics[width=\linewidth]{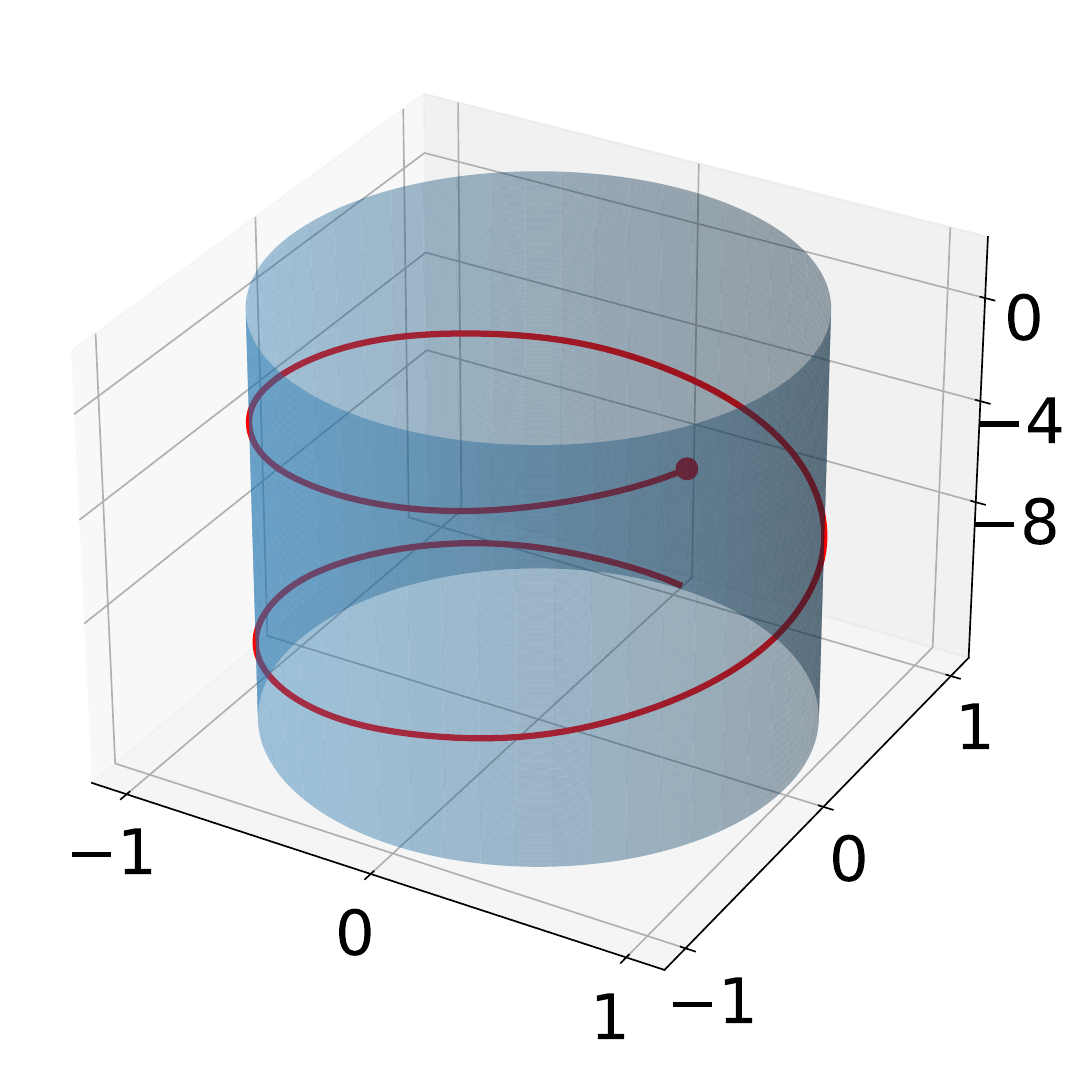}
  \caption{$\beta = \frac{\pi}{3}, \alpha_I = \frac{\pi}{3}$}

\end{subfigure} 
\begin{subfigure}{0.24\textwidth}
\includegraphics[width=\linewidth]{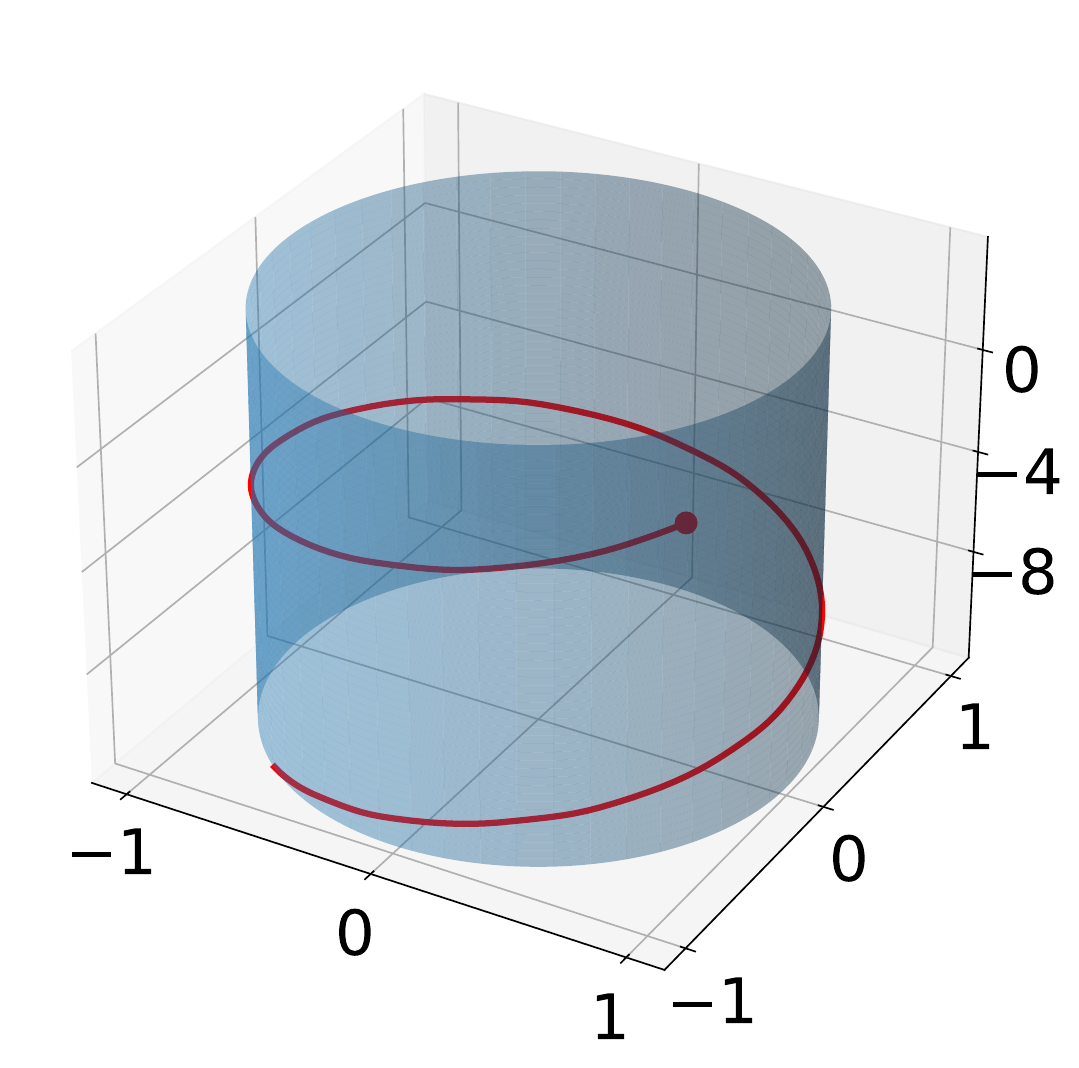}
  
\caption{$\beta = \frac{\pi}{3}, \alpha_I = \frac{\pi}{2}$}
  \end{subfigure}
  \begin{subfigure}{0.24\textwidth}
\includegraphics[width=\linewidth]{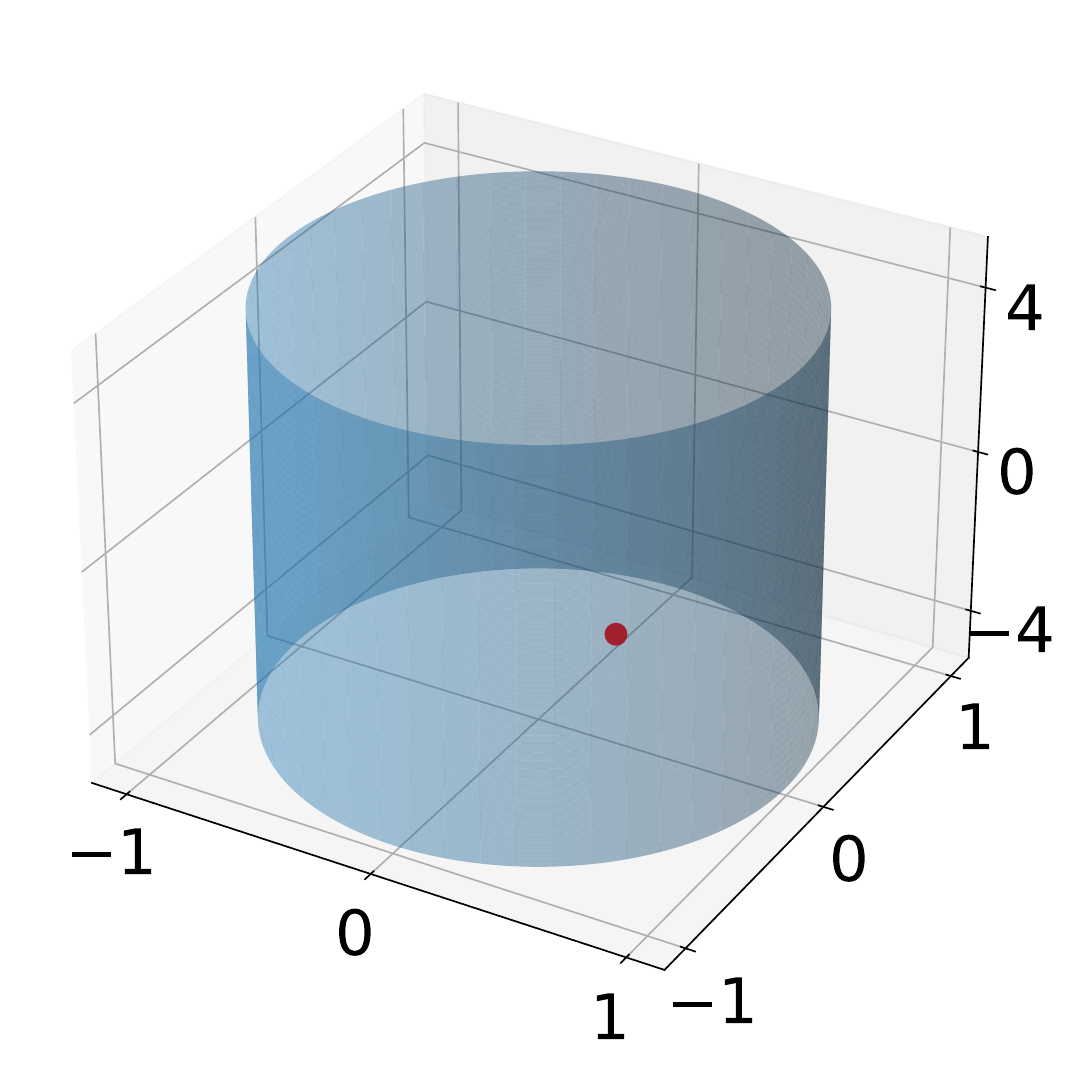}
  \caption{$\beta = \frac{\pi}{2}, \alpha_I = 0$}

\end{subfigure} 
\begin{subfigure}{0.24\textwidth}
  \includegraphics[width=\linewidth]{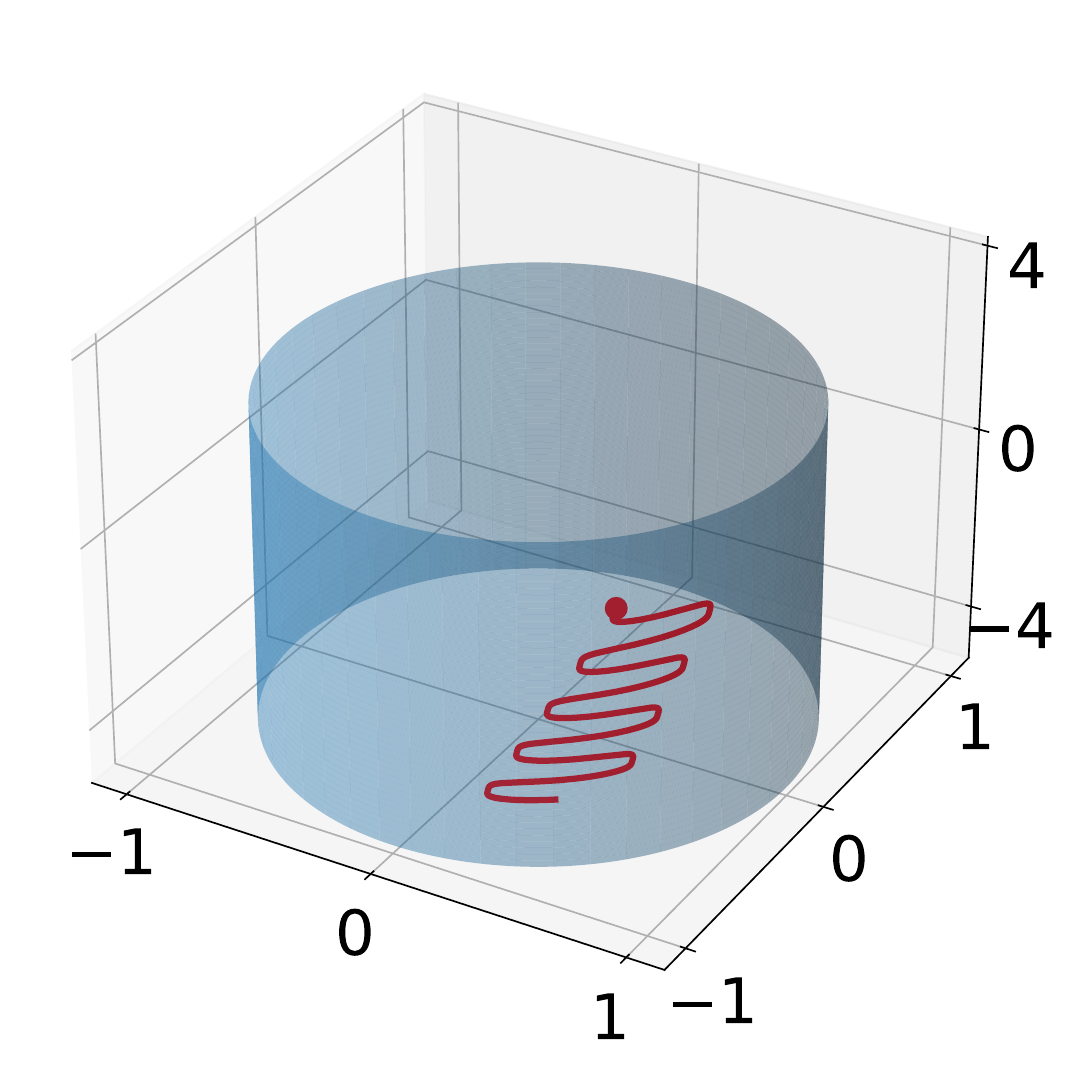}
  
 \caption{$\beta = \frac{\pi}{2}, \alpha_I = \frac{\pi}{6}$ (Case D)}
  \label{matrix_d}
\end{subfigure}
\begin{subfigure}{0.24\textwidth}
\includegraphics[width=\linewidth]{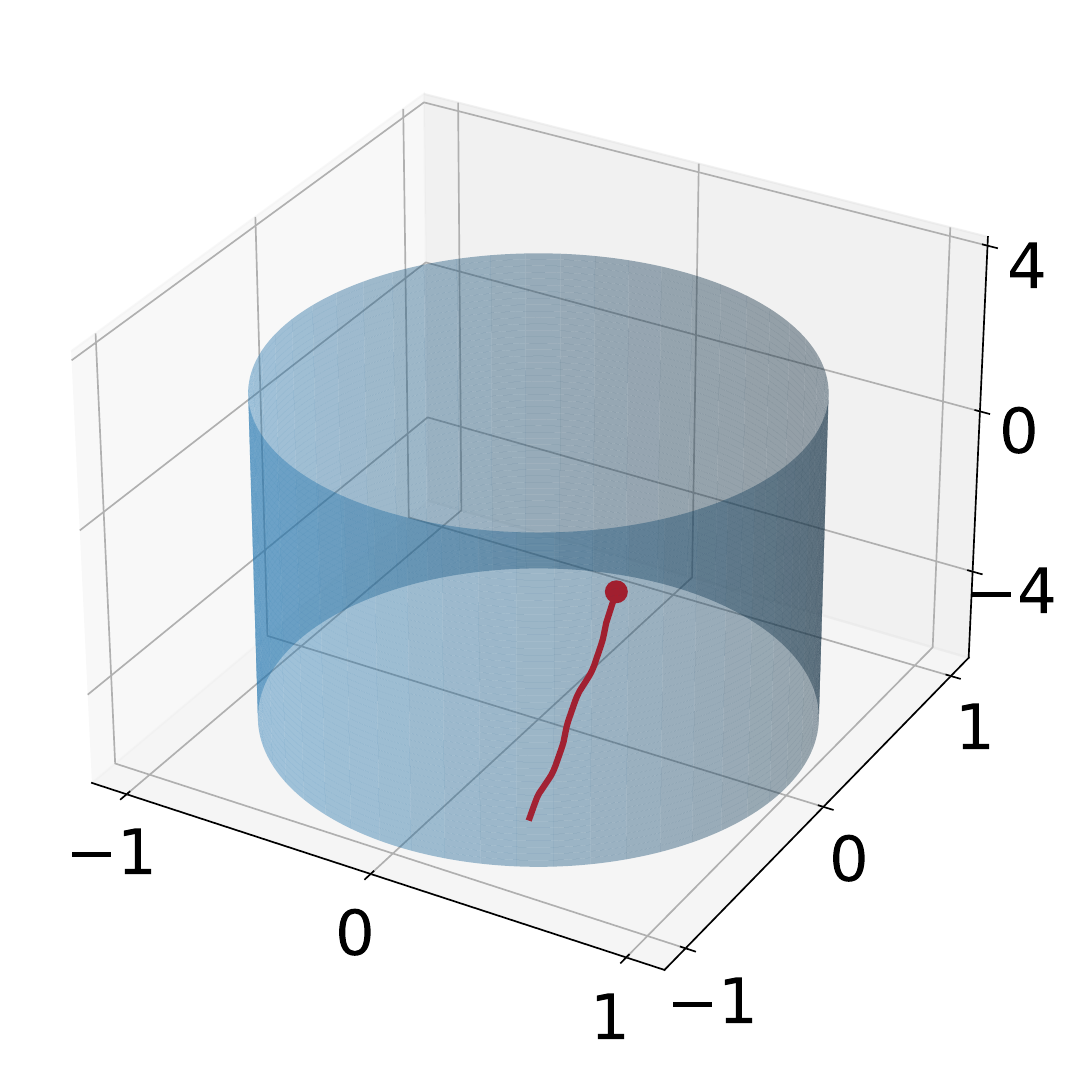}
   \caption{$\beta = \frac{\pi}{2}, \alpha_I = \frac{\pi}{3}$}

\end{subfigure} 
\begin{subfigure}{0.24\textwidth}
  \includegraphics[width=\linewidth]{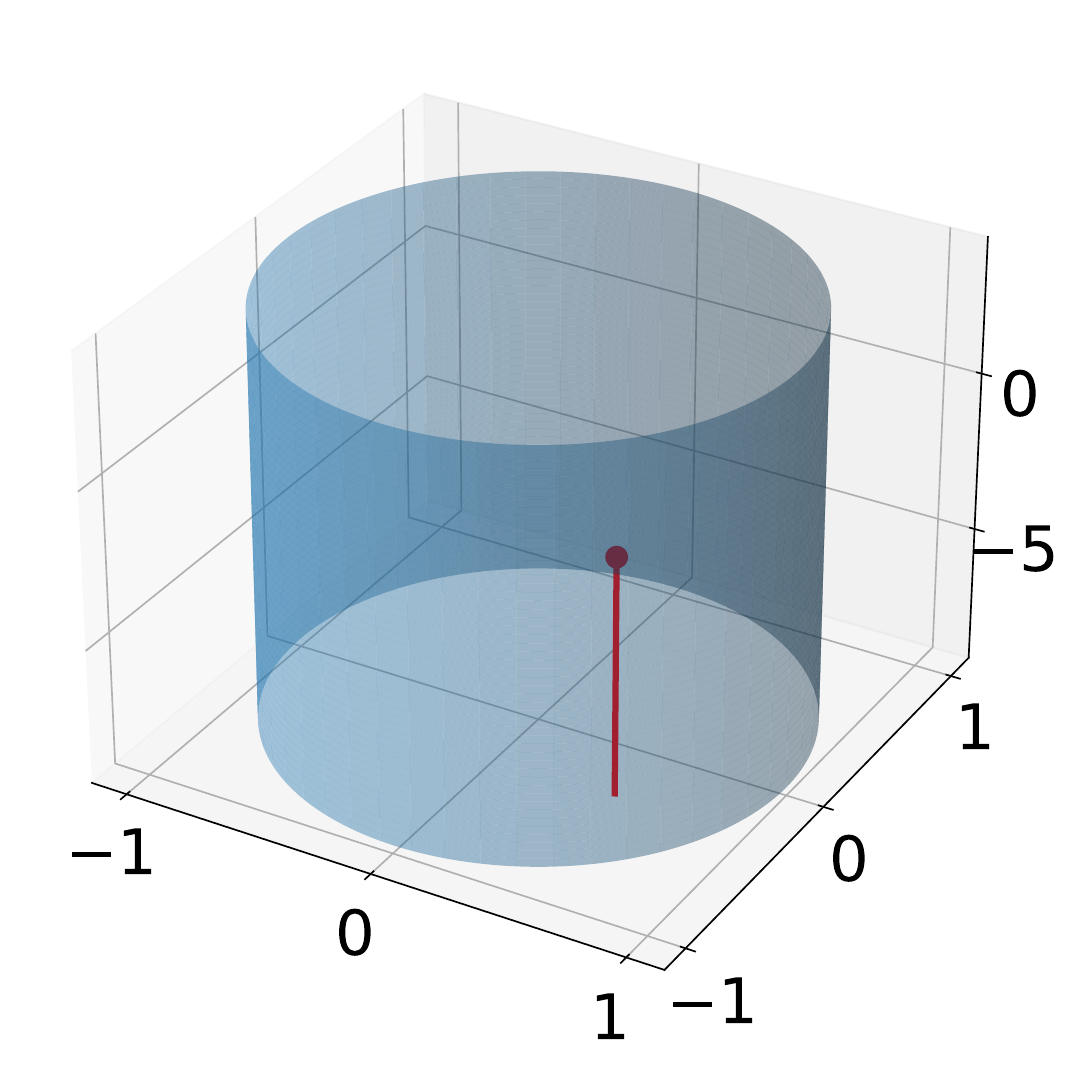}
   \caption{$\beta = \frac{\pi}{2}, \alpha_I = \frac{\pi}{2}$ (Case B)}
 
  \label{matrix_b}
\end{subfigure}

\caption{Dipole trajectories for the two-dipole system for different initial locations denoted by $\beta$ and relative orientation $\alpha_I$, (a) $\beta = 0, \alpha_I = 0$, (b) $\beta = 0, \alpha_I = \frac{\pi}{6}$ (Case C), (c) $\beta = 0, \alpha_I = \frac{\pi}{3}$, (d) $\beta = 0, \alpha_I = \frac{\pi}{2}$ (Case A), (e)  $\beta = \frac{\pi}{6}, \alpha_I = 0$,  (f)  $\beta = \frac{\pi}{6}, \alpha_I = \frac{\pi}{6}$, (g)  $\beta = \frac{\pi}{6}, \alpha_I = \frac{\pi}{3}$, (h)  $\beta = \frac{\pi}{6}, \alpha_I = \frac{\pi}{2}$ (Case E), (i)  $\beta = \frac{\pi}{3}, \alpha_I = 0$, (j)  $\beta = \frac{\pi}{3}, \alpha_I = \frac{\pi}{6}$, (k) $\beta = \frac{\pi}{3}, \alpha_I = \frac{\pi}{3}$, (l) $\beta = \frac{\pi}{3}, \alpha_I = \frac{\pi}{2}$, (m)  $\beta = \frac{\pi}{2}, \alpha_I = 0$, (n) $\beta = \frac{\pi}{2}, \alpha_I = \frac{\pi}{6}$ (Case D), (o) $\beta = \frac{\pi}{2}, \alpha_I = \frac{\pi}{3}$, (p)  $\beta = \frac{\pi}{2}, \alpha_I = \frac{\pi}{2}$ (Case B). The special cases A-E of the main text are marked in parenthesis.   } 
\label{matrix}
\end{figure}

In this section, we present a few additional plots that will be a useful addition to the figures already presented in the main text. In Fig.~(\ref{matrix}) we present a matrix map showing the dipole trajectories on the cylinder surface as the initial location of the dipoles and their relative orientation $\alpha_I$ are varied. For this purpose, we choose one of the dipoles to be situated at the origin at $t=0$ oriented parallel to the $\hat{\theta}$ direction. Next, we define a parameter $\beta$ which denotes the angle made by the position vector of the initial location of the second dipole from the $\hat{\theta}$ direction. We show the resultant dipole trajectories as the parameters $\alpha_I$ and $\beta$ are varied in steps of $\frac{\pi}{6}$ from 0 to $\frac{\pi}{2}$, resulting in a ($4\times 4$) matrix of 3d plots of trajectories on the cylinder. In this matrix of figures, we identify the special cases A-E of Fig.~(\ref{3d}) of the main text, arising for specific choices of the two parameters $\beta$ and $\alpha_I$. Note that Case E has  $\beta= \pi/4$ for the plot in the main text but generates a qualitatively similar trajectory for $\beta=\pi/6$; hence we identified this point in parameter space with Case E. In Fig.~(\ref{azoomedfig}) we provide the temporal evolution of orientations of the dipole pair and the respective trajectories in $\theta-z$ plane, in addition to the plots presented in Fig.~(\ref{zoomedfig}) of main text.

\newpage
\begin{figure}[H]
\hspace{-2cm}
\begin{subfigure}{0.65\textwidth}
\includegraphics[width=\linewidth]{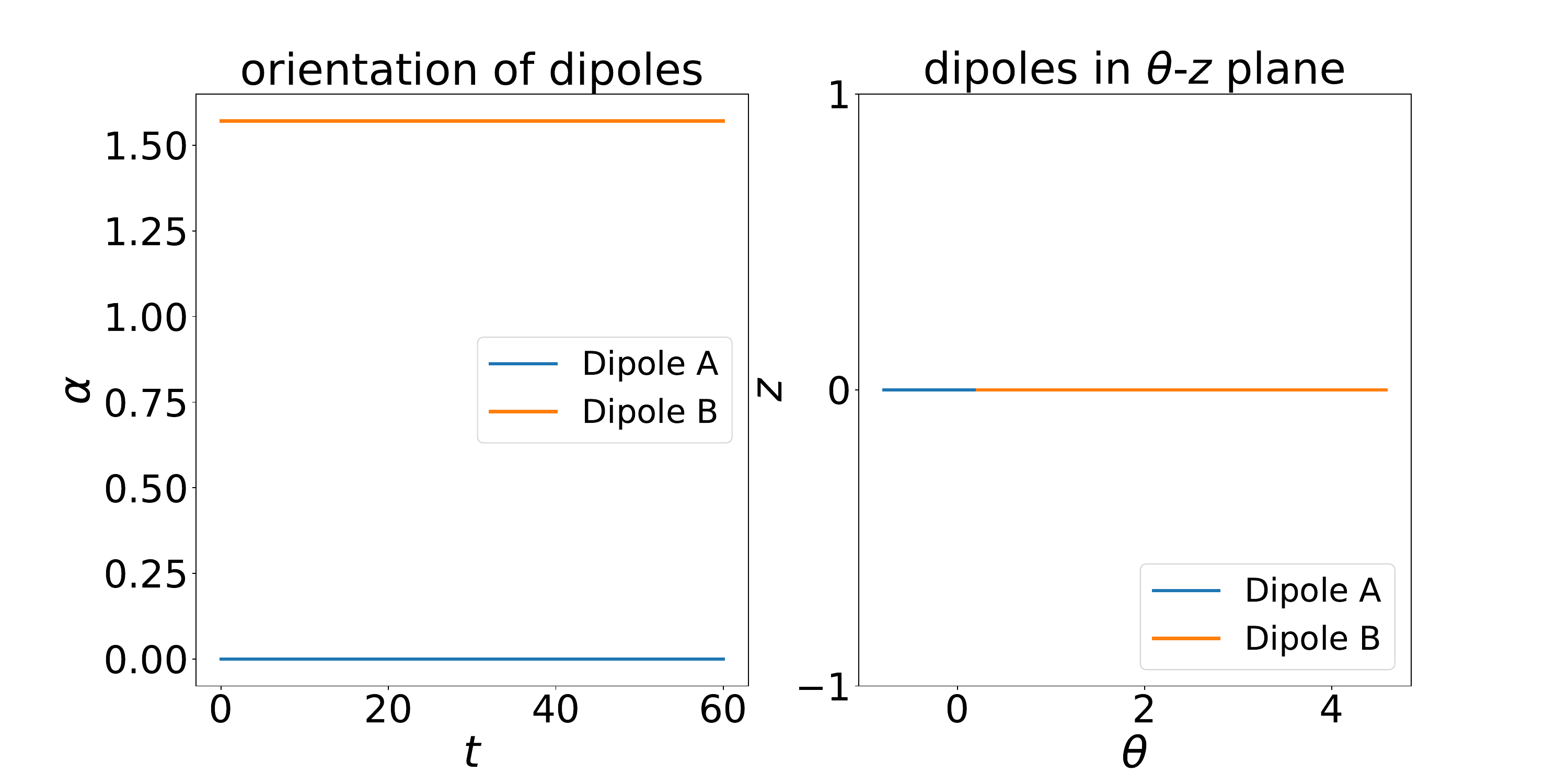}
  
  \subcaption{Case A}

\end{subfigure} \hspace{-1.7cm}
\begin{subfigure}{0.65\textwidth}
  \includegraphics[width=\linewidth]{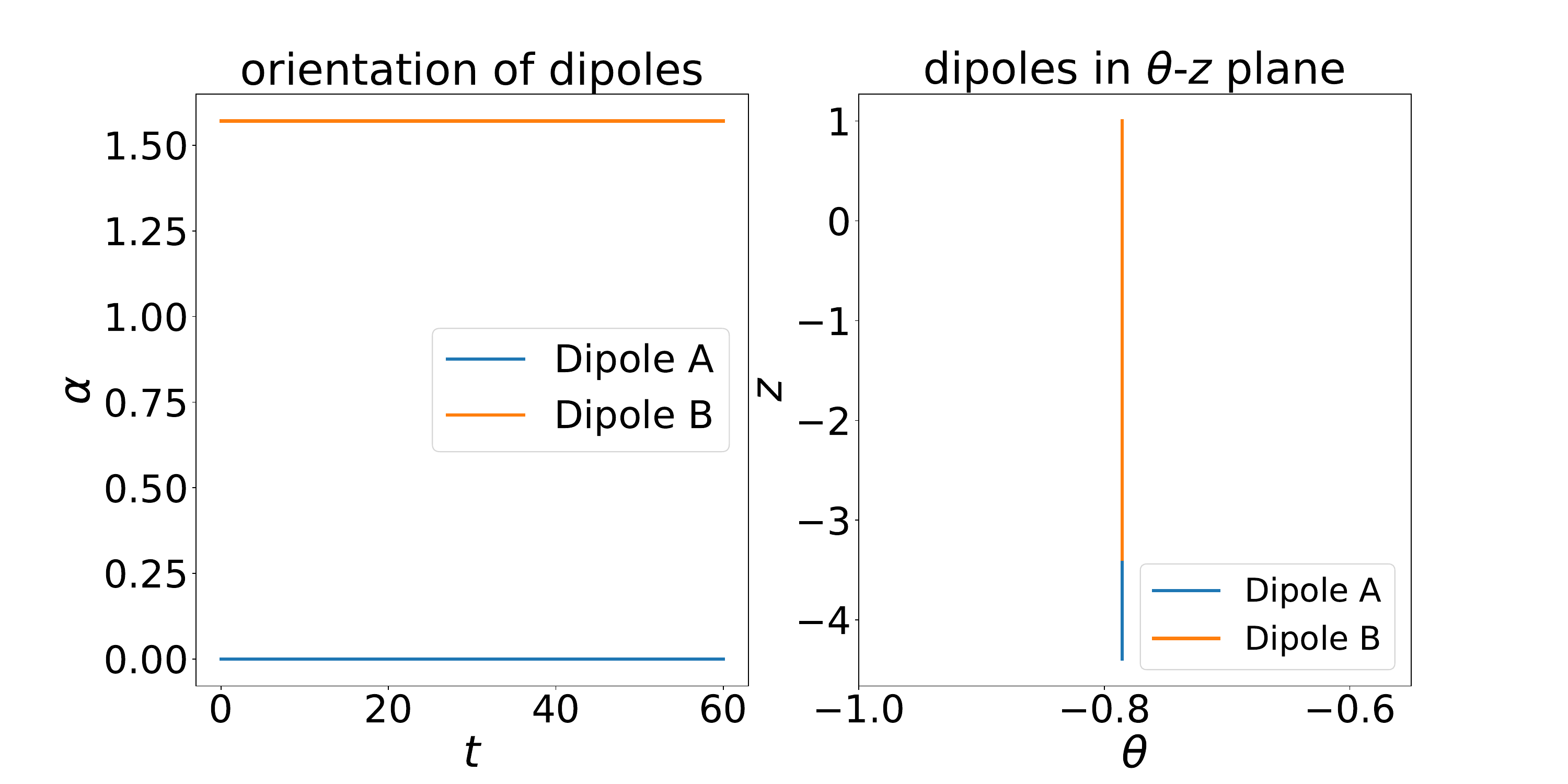}
  
  \subcaption{Case B}

\end{subfigure}
\end{figure}
\begin{figure}[H]\ContinuedFloat
 \hspace{-2cm}
\begin{subfigure}{0.65\textwidth}
\includegraphics[width=\linewidth]{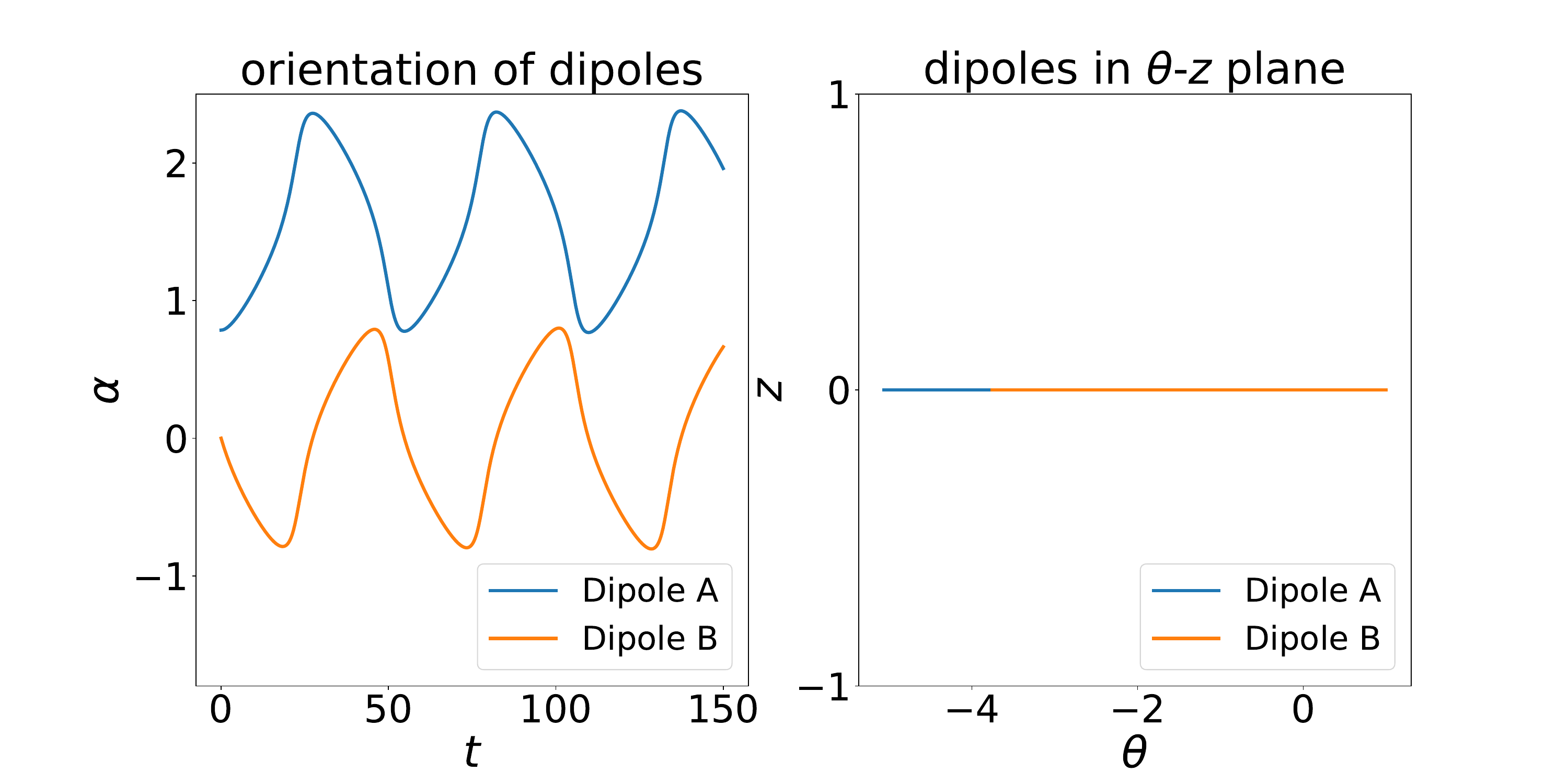}
  
  \subcaption{Case C}

\end{subfigure} \hspace{-1.7cm}
\begin{subfigure}{0.65\textwidth}
  \includegraphics[width=\linewidth]{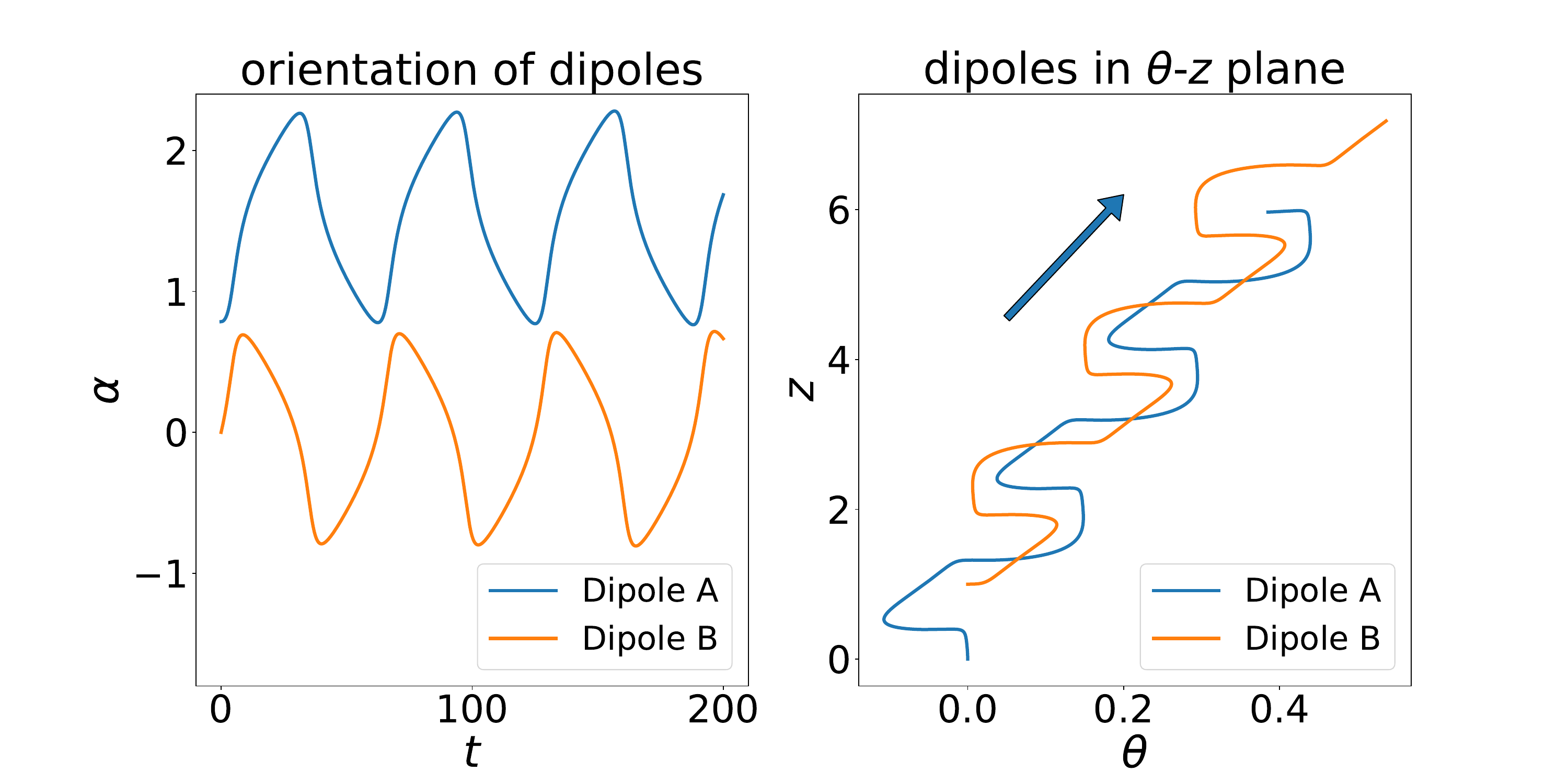}
  
  \subcaption{Case D}

\end{subfigure}
\end{figure}
\hspace{-2cm}
\begin{figure}[H]\ContinuedFloat
\hspace{-2cm}
  \begin{subfigure}{0.65\textwidth}
  \includegraphics[width=\linewidth]{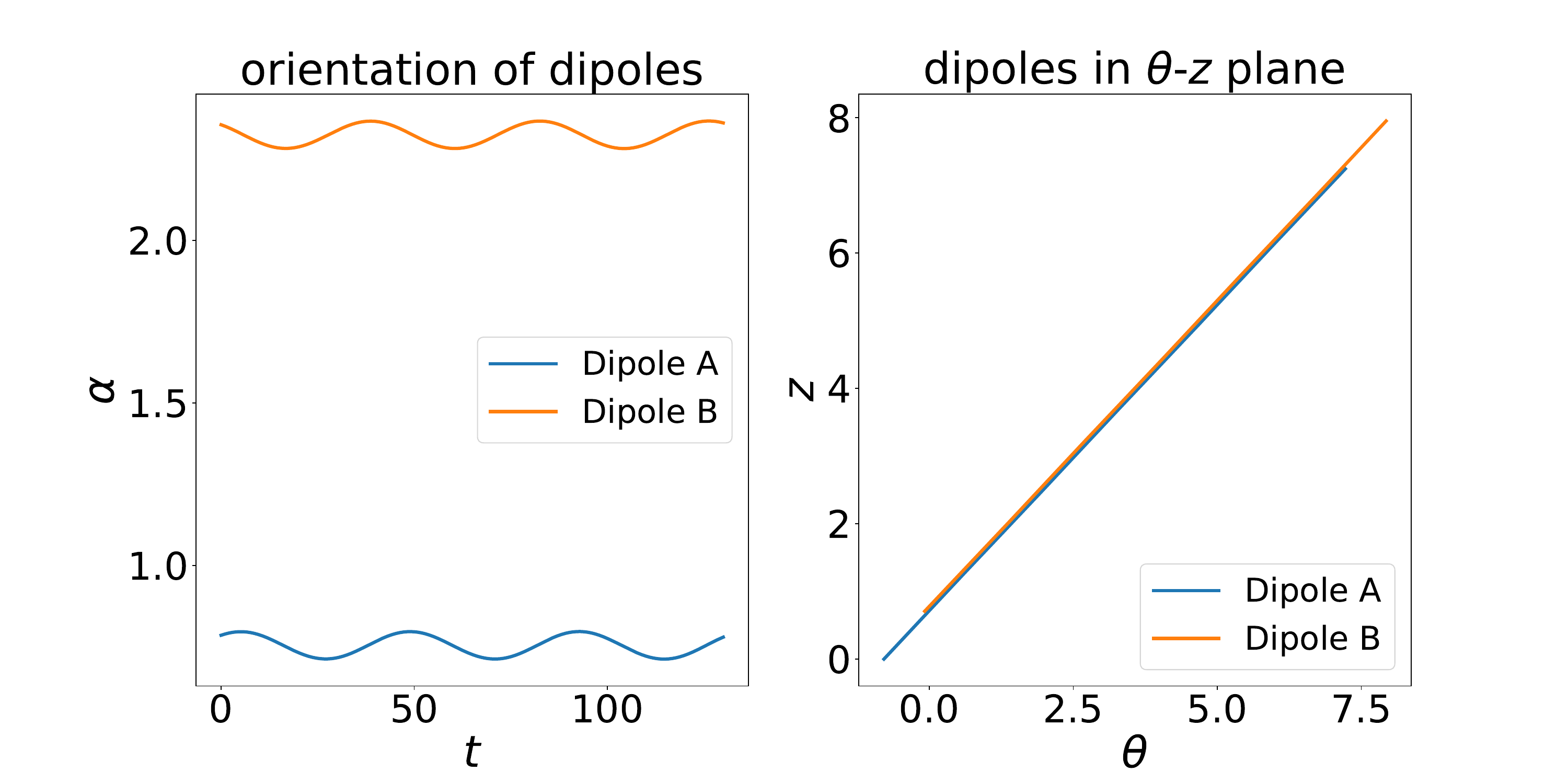}
  \subcaption{Case E}

\end{subfigure} 
\hspace{-1.7cm}
\begin{subfigure}{0.65\textwidth}
  \includegraphics[width=\linewidth]{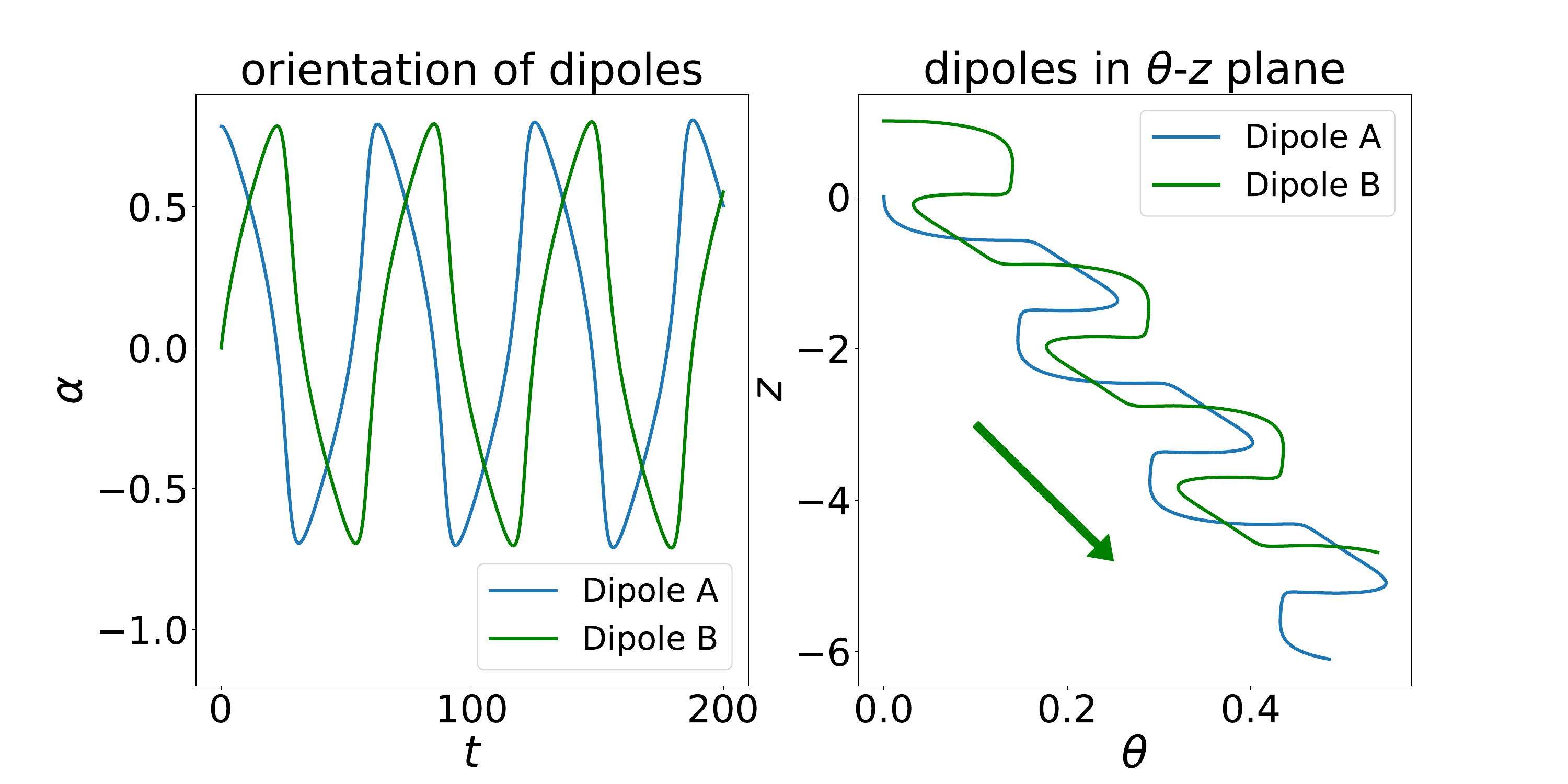}
  
  \subcaption{Case F}
  
\end{subfigure}
\caption{\label{azoomedfig} The temporal evolution of orientations of the dipole pair and the respective trajectories in $\theta-z$ plane in addition to the plots presented in Fig.~(\ref{zoomedfig}) of main text. The initial conditions and parameters are the same as those of main text. For case F, the green-colored trajectory is for the dipole with strength $\kappa=-1$.}

\end{figure}
\newpage 
\textbf{Data Availability} The analytical data that support the findings of this study are
available within the article and its supplementary material. Numerical
details and additional data are available from the authors upon reasonable
request.

\begin{footnotesize}

\end{footnotesize}

\end{document}